\documentclass[]{aa}  

\usepackage{graphicx}
\usepackage{txfonts}
\usepackage{placeins}
\usepackage[options]{xcolor}

\usepackage{natbib}
\bibpunct{(}{)}{;}{a}{}{,} 

\begin{document} 

\title{An LBT view of the co-rotating group of galaxies around NGC~2750:\\Deep imaging and new satellite candidates}
\subtitle{}
	
   \author{
        S.~Taibi\inst{1,2}
        \and
        M.~S.~Pawlowski\inst{2}
        \and
        O.~Müller\inst{1,3,4}
        \and
        M.~Bílek\inst{5}
        \and
        M.~P.~Júlio\inst{2,6}
        \and
        K.~J.~Kanehisa\inst{2,6}
        \and
        M.~Jovanovi{\'c}\inst{5}
        \and
        A.~Lalovi{\'c}\inst{5}
        \and
        S.~Samurovi{\'c}\inst{5}
        }

   \institute{
        Institute of Physics, Laboratory of Astrophysics, École Polytechnique Fédérale de Lausanne (EPFL), 1290 Sauverny, Switzerland\\
        \email{salvatore.taibi@epfl.ch}
    \and
        Leibniz-Institut für Astrophysik Potsdam (AIP), An der Sternwarte 16, 14482 Potsdam, Germany
    \and
        Institute of Astronomy, Madingley Rd, Cambridge CB3 0HA, UK
    \and
        Visiting Fellow, Clare Hall, University of Cambridge, Cambridge, UK
    \and
        Astronomical Observatory, Volgina 7, 11060 Belgrade, Serbia
    \and
        Institut für Physik und Astronomie, Universität Potsdam, Karl-Liebknecht-Straße 24/25, 14476 Potsdam, Germany
        }

\date{Received; accepted}

 
\abstract
   {Some galaxies such as the Milky Way and Andromeda display coherently rotating satellite planes, posing tensions with cosmological simulations. NGC~2750 has emerged as an additional candidate system hosting a co-rotating group of galaxies.}
   {This work aims to verify the presence of a coherent plane of satellites around NGC~2750 by identifying new candidate dwarf galaxies and low surface brightness features.}
   {We conducted deep photometric observations of NGC~2750 and its surroundings over an area of $\sim35\arcmin\times30\arcmin$ using the Large Binocular Telescope (LBT) in the \textit{g}- and \textit{r}-bands. Standard data reduction techniques were employed to improve the detection of low-surface-brightness-features down to a depth of $\mu_r\sim31$~mag\,arcsec$^{-2}$. We analysed the morphology of NGC~2750 and other galaxies in this system for signs of tidal interactions and performed structural and photometric analyses of recently identified candidate satellites.}
   {Our observations led to the discovery of six new candidate dwarf galaxies, one of which exhibits properties consistent with an ultra-diffuse galaxy. We identified tidal features around NGC~2750, suggesting past interactions with its neighbouring satellites. While the spatial distribution of satellites suggests a moderate flattening, and this flattening is supported by the new candidates, follow-up spectroscopic measurements of the new candidates have the potential to bolster or diminish the strong kinematic coherence observed previously. The luminosity function of NGC~2750 shows an excess of bright satellites compared to similar systems, contributing to the growing evidence of discrepancies between observed satellite distributions and predictions from cosmological simulations.}
   {}


\maketitle


\section{Introduction}
\label{sec:intro}

The $\Lambda$ Cold Dark Matter ($\Lambda$CDM) cosmological model has been very successful in explaining the large-scale structure and evolution of the Universe \citep{Tempel2014,Planck2016}. However, comparing the results of cosmological simulations with observations has revealed a number of tensions, especially at the low-mass scales of dwarf galaxies \citep*{Bullock+Boylan-Kolchin2017,Sales2022}. 

A serious challenge has been posed by the existence of planes of satellite systems, such as those found around the Milky Way \citep{Kroupa2005}, M31 \citep{Conn2012,Ibata2013}, Centaurus A \citep{Tully2015,Muller2018}, and at larger distances NGC~4490 \citep{Pawlowski2024}. These features are highly flattened and kinematically coherent, which is in clear contrast to the more random distributions obtained from cosmological simulations \citep[see][for a review]{Pawlowski2021}.

The plane of satellites problem seems to pose a stronger tension than others involving dwarf galaxies because it is mostly independent of the internal properties of individual systems. It offers the opportunity to critically test the cosmological model without being dominated by the details of how baryonic effects are implemented in simulations or whether alternative types of dark matter are considered \citep{Pawlowski2021,Sales2022}. 

Several mechanisms for the formation of satellite planes have been proposed. Some of them involve interactions between galaxies, such as mergers and flybys of neighbouring galaxies \citep[e.g.][]{Pawlowski2011,Smith2016,Bilek2018}, although major mergers seem to play a marginal role if considered in a full $\Lambda$CDM context \citep{Kanehisa2023}. 
Other formation mechanisms take into account the group infall of dwarf galaxies \citep[e.g.][]{Li+Helmi2008,Garavito-Camargo2021,Vasiliev2021,Julio2024} or the accretion along filaments of the cosmic web \citep[e.g.][]{Libeskind2015}. 

To date, none of the proposed mechanisms seems to satisfactorily explain the formation of satellite planes (see \citealp{Pawlowski2018,Pawlowski2021}, but also \citealp{Taibi2024}).
Therefore, it is essential from an observational point of view to expand the number of known systems hosting coherent planar structures and to better characterise their properties in order to distinguish between the different formation scenarios.

Multiple galaxies within 10 Mpc of us have been reported to host candidate satellite planes, including M~81 \citep{Chiboucas2013,Muller2024a}, M~101 \citep{Muller2017}, NGC~253 (\citealp{Martinez-Delgado2021}, but see \citealp{Mutlu-Pakdil2024}), as well as several systems from the MATLAS survey \citep{Heesters2021}. However, they generally lack velocity measurements to establish the degree of kinematic coherence. 

An interesting candidate plane with a confirmed co-rotating group of dwarf galaxies has been reported by \citet{Paudel2021} around the spiral galaxy NGC~2750. They found that its six satellites, which have known radial velocities, appear to orbit the galaxy in the same direction and that their distribution across the sky is elongated. The authors calculated a high correlation between the satellites' relative line-of-sight velocities and their sky-projected distances from the host, which resembles a small-scale version of Cen~A. 

Despite a poorly constrained distance estimate \citep{Bottinelli1985,Tully+Fisher1988,Sorce2014}, NGC~2750 appears to reside in a rarefied environment. Assuming a Hubble flow distance of 42~Mpc, there are only two other galaxies, NGC~2743 and NGC~2735, in a volume of 1~Mpc$^3$ that show similar luminosity and redshift to NGC~2750. Their projected separations of $>500$~kpc should rule out any possible association, but it is not excluded that they could be part of a loose group of galaxies \citep{Paudel2021}. 

The combination of phase-space properties and the relative isolation of the NGC~2750 system motivated us to carry out an observing campaign to obtain wide-area deep photometric images of its surroundings. 
Theories suggest that satellite planes can form after mergers or flybys of massive galaxies \citep{Smith2016,Bilek2018}. We might then expect such interactions to leave characteristic traces around the central hosts, namely tidal features and asymmetric massive stellar halos. Such structures tend to have very low surface brightness (typically $\mu_r>28$~mag\,arcsec$^{-2}$; \citealp{Bilek2020}) and require dedicated observations to detect them. 

Deep images also offer the possibility of discovering new candidate satellites in the host's vicinity, similar to what has been done in other systems of the local Universe \citep{Mao2021,Carlsten2022,Crosby2024,Zaritsky2024}, thus providing an additional opportunity to test the planarity of the NGC~2750 system. In particular, the Large Binocular Telescope (LBT) has proved to be an ideal facility for detecting candidate dwarf galaxies, as shown in surveys such as SSH \citep{Annibali2020}, LBT-SONG \citep{Garling2021}, and LIGHTS \citep{Trujillo2021,Zaritsky2024}.

This article is structured as follows. In Sect.~\ref{sec:data}, we describe the acquisition and reduction of the observational dataset, including its photometric calibration. In Sect.~\ref{sec:distance} we discuss the distance of the group.
Section \ref{sec:analysis} is dedicated to the analysis of the low surface brightness features around NGC~2750 and the discovery of new candidate satellites in the field of view. In Sect.~\ref{sec:discussion}, we discuss the physical properties of the dwarf galaxy candidates compared to other known systems, to then study the luminosity function of NGC~2750 and the group coherent motion in light of the new satellite candidates. Finally in Sect.~\ref{sec:end} we present our conclusions.

\section{Data}
\label{sec:data}

The observations were acquired with the LBT on Mount Graham observatory using the Large Binocular Cameras (LBC) in dual mode \citep{Giallongo2008,Speziali2008}. The LBC are two wide-field imagers optimised for acquisitions, respectively, in the blue (from $\sim$3500~\AA~to 6500~\AA) and red (from $\sim$5500~\AA~to 1~$\mu$m) part of the visible. At the focal plane, each LBC is composed of four $2{\rm k}\times4{\rm k}$ CCDs covering a field of view (FoV) of $\sim 25\times 23$~arcmin$^2$ with a pixel scale of $\sim0.225$~arcsec\,pix$^{-1}$.

The images were obtained during the night of 10 April 2023 under equivalent dark conditions using the \textit{g}- and \textit{r}-band SLOAN filters in the blue and red arms, respectively.\footnote{LBTB programme: AIP-2022B-002 (PI: S.~Taibi).}
Night conditions were good, with a clear sky, low humidity and a seeing ranging from 0.8 to 1.2 arcsec. The total telescope time was approximately 2~hrs, which accounted for 1.5~hrs on source in both pass-bands plus 0.5~hrs of overheads. The time on source was split into 30 dithered exposures of 180~s each.

The observing strategy was based on that followed by \citet{Trujillo2021}. In summary,  the idea was to obtain a background as homogeneous as possible by using the scientific data itself as a flat-field \citep[see also][]{Trujillo+Fliri2016}. This required the dithered exposures to follow a path with step size of the order of the main target dimensions. In our case, NGC~2750 has a diameter of $D_{25}\sim2$~arcmin \citep{Paturel2000}, so we used steps of $2.5$~arcmin. In this way, the central host was located in different regions of the LBC chips, allowing for optimal flat-field correction, since we also covered the spaces between the chips. As in \citet{Trujillo2021}, we did not rotate the cameras, but instead we fixed all exposures to a single position angle.

\subsection{Data reduction and image co-addition}

We followed a standard data reduction procedure, with appropriate variations when necessary.
All steps are to be considered per chip and filter, unless stated otherwise. 
The reduction pipeline made large use of \texttt{ccdproc} \citep{ccdproc}, an image reduction package, together with \texttt{photutils} \citep{photutils} and other \texttt{Astropy} \citep{Astropy2022} routines.

For the bias correction we first subtracted the overscan from all the calibration and science frames. We then combined 15 bias images to create a masterbias using \texttt{ccdproc.combine} with a 3$\sigma$ clipping median stacking. The masterbias was then subtracted from the rest of the frames to remove any residual systematic pattern.

For the flat-field correction, we followed the procedure highlighted again in \citet{Trujillo2021}. We first normalised the science frames by defining a normalisation ring that crossed the four CCDs. This normalisation ring was centred on the pixel (1024, 2998) of the central CCD and had an inner radius of 2505 pixels and a width of 200 pixels.
We took the median value within the ring and divided it to the flux in each CCD of the science frames. We then combined the scientific frames to create a masterflat using again \texttt{ccdproc.combine} with a 3$\sigma$ clipping median. Finally, each individual scientific frame was divided by the masterflat. 

The last step of the data reduction involved the subtraction of the sky background. For this task, we masked all the sources above 2$\sigma$ of the noise background using the \textit{detect\_sources} function of \texttt{photutils.segmentation}. In this way, we created a sky image of which we calculated the median value, which we then subtracted from all the science frames.

Finally, in order to co-add all the individual exposures, it was necessary to carry out their astrometry. We used Astrometry.net \citep[v.0.93;][]{Lang2010} to calculate the astrometric solution of the individual science frames, taking Gaia eDR3 \citep{Gaia2021} as the reference catalogue. We then created catalogues of detected sources generated with SExtractor \citep[v.2.25;][]{Bertin+Arnouts1996} and calculated the image distortion coefficients using SCAMP \citep[v.2.0.4;][]{Bertin2006}. As a final step, we ran SWarp \citep[v.2.38;][]{Bertin2002} to median combine the individual images by placing them on a common grid of $9500\times9200$~pixels, using a LANCZOS3 resampling.

\subsection{Photometric calibration}

\begin{figure*}
    \centering
    \includegraphics[width=.95\textwidth]{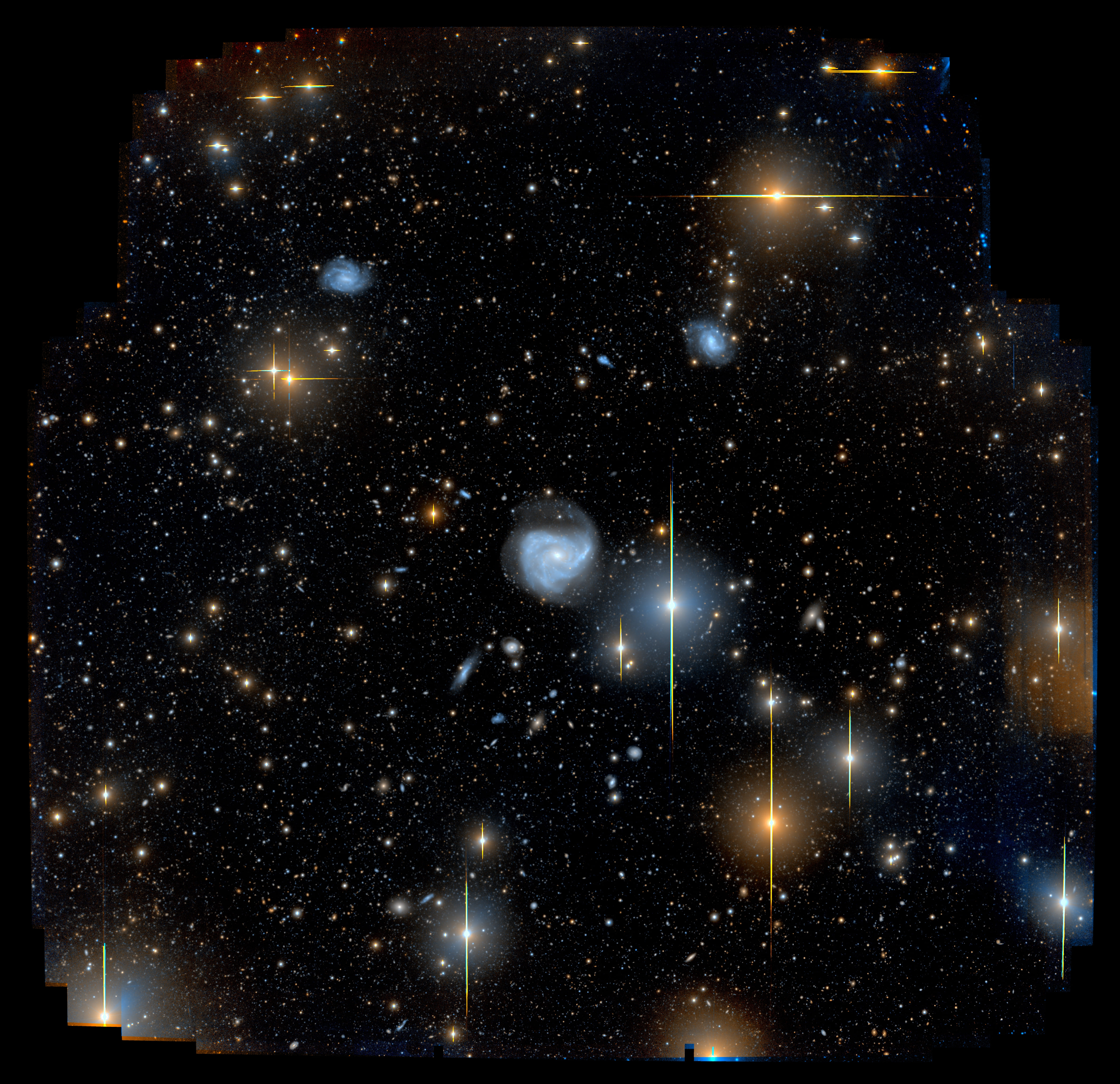}
    \caption{Colour composition of the surveyed area around NGC~2750 obtained by combining the final co-added images in the \textit{g}- and \textit{r}-bands with an average \textit{(g+r)/2} image. The image FoV is approximately $35\arcmin\times30\arcmin$. North is up and east to the left.}
    \label{fig:color}
\end{figure*}

We performed the photometric calibration of our data using an SDSS DR18 catalogue \citep{Almeida2023} covering the same area.
We generated the catalogue querying only for stars with clean photometry having magnitudes between 18 and 22 mag in both \textit{g}- and \textit{r}-band filters. Sources in our data instead were obtained using SExtractor. 
We found about 300 stars in common between SDSS and our data, using their Petrosian magnitudes to perform the photometric calibration. 

For our \textit{g}- and \textit{r}-band co-added images, we calculated only the zeropoint, finding no need to add a colour term between these and SDSS. We retrieved the following values: ZP$_g = 24.621 \pm 0.011 \pm 0.01$ and ZP$_r = 23.771 \pm 0.008 \pm 0.01$ mag (where the first error term is the statistical error, while the second term is the typical SDSS zero-point photometric error; \citealp{Ivezic2004}). 

To estimate the surface brightness limit we measured the standard deviation in 10 boxes of the size of 10$\times$10\,arcsec$^2$. This gave a 3$\sigma$ surface brightness limit of 30.2\,mag/arcsec$^2$ for the \textit{g}-band and of 30.6\,mag/arcsec$^2$ for the \textit{r}-band. The depth limit is comparable to that obtained by \citet{Trujillo2021}.

A colour composition obtained by combining the final co-added images in \textit{g} and \textit{r} with an average image \textit{(g+r)/2} is shown in Fig.~\ref{fig:color} (using \texttt{astscript-color-faint-gray} from \textit{Gnuastro}, \citealp{Infante-Sainz+Akhlaghi2024}).

\section{Distance to the NGC~2750 system}
\label{sec:distance}

\begin{table*}
    \caption{Observed properties and Tully-Fisher distance estimates of the known satellite galaxies in the surveyed region around NGC~2750.}
    \label{table:0}
    \centering          
    \begin{tabular}{c c c c c c c c c c c}    
      \hline\hline
      ID & Name & R.A.  & Dec.  & $r_0$ & $(g-r)$ & P.A.  & $b/a$ & $q_0$ & $W_{50}$       & $D_{\rm TF}$ \\ 
         &      & (deg) & (deg) & (mag) &  (mag)  & (deg) &       &       & (km\,s$^{-1}$) & (Mpc)    \\
      \hline  
      D1 & UGC~4774     & $136.57416$ & $+25.58371$ & $15.10\pm0.04$ & 0.20 & $78$  & $0.73$ & 0.2  & $142\pm2$ & $50\pm3$ \\ 
      D2 & UGC~4764     & $136.35966$ & $+25.55069$ & $14.81\pm0.08$ & 0.10 & $28$  & $0.71$ & 0.2  & $148\pm2$ & $44\pm3$ \\
      D3 & LEDA~213565  & $136.42281$ & $+25.54056$ & $17.62\pm0.12$ & 0.01 & $62$  & $0.64$ & 0.35 & $ 68\pm6$ & $36\pm7$ \\
      D4 & LEDA~1737674 & $136.50380$ & $+25.46980$ & $16.89\pm0.02$ & 0.10 & $60$  & $0.39$ & 0.35 & $ 61\pm4$ & $16\pm2$ \\
      D6 & LEDA~2807197 & $136.48517$ & $+25.35100$ & $18.42\pm0.03$ & 0.10 & $105$ & $0.62$ & 0.5  & $ 75\pm4$ & $53\pm8$ \\
      \hline                
    \end{tabular}  
    \tablefoot{The quantities listed in columns are: (1-2) galaxy's ID as in \citet{Paudel2021} and name, (3-8) central coordinates, de-reddened \textit{r}-band magnitudes and $(g-r)$ colours, position angles (measured from north to east, with $\pm2^\circ$ errors) and axis ratios (with $\pm0.02$ errors) obtained in this work, (9) intrinsic axial ratio following \citet{Roychowdhury2013}, (10) observed HI velocity widths at 50\% of peak from \citet{Haynes2018}, and (11) estimated Tully-Fisher distances using the calibration from \citet{Kourkchi2020}.}
\end{table*}

The distance estimate in the literature for NGC~2750 is rather uncertain. Values derived from the Tully-Fisher relation average $\sim20$~Mpc, with values ranging from 9~Mpc to 39~Mpc \citep{Bottinelli1985,Tully+Fisher1988,Sorce2014}. Such a large scatter is mostly due to the poorly constrained inclination of NGC~2750, which is almost face-on. 
On the other hand, the Hubble flow distance predicted from the Cosmicflows-4 distance-velocity calculator \citep{Valade2024} is of 42~Mpc \citep[as also in][]{Paudel2021,Tully2023}.

Our initial intention to improve on the current state of affairs was to model the inclination of the outer disk of NGC~2750 to re-measure its Tully-Fisher distance. The axis ratio and position angle are typically stable with the radius, following the method outlined in \citet{Monelli+Trujillo2019}, and with deeper observations, a better constraint on the inclination seemed plausible.

We modelled NGC~2750 using the \textit{Ellipse} function of \texttt{photutils.isophote} after masking the bright objects around it, for which we used the \textit{detect\_sources} function of \texttt{photutils.segmentation}, assuming a 3$\sigma$ threshold. The radial axis-ratio and position angle profiles obtained from the elliptical modelling did not attain a constant value once the periphery was reached, due to the tidal features that drive the elongation of the galaxy (see Sect.~\ref{sec:analysis}). Therefore, the distance value of NGC~2750 remains uncertain.

Instead of measuring the distance for NGC~2750 itself, we modelled its satellite galaxies with HI emission to get an overall estimate of the Tully-Fisher distance of the group. Applying the same methodology as described in the last paragraph, we measured the axis ratio $b/a$ of galaxies D1 to D6 (following \citealp{Paudel2021} nomenclature), which all exhibit HI emission and have measured $W_{50}$ values \citep{Haynes2018}.\footnote{$W_{50}$ is defined as the velocity width of the HI line profile measured at 50\% peak level.} 
Their inclination was derived using the following equation: ${\rm cos}\,i = [((b/a)^2-q_0^2)/(1-q_0^2)]^{1/2}$, where the intrinsic axial ratio $q_0$ depends on the morphological type and brightness of the system (following \citealp{Roychowdhury2013}, see Table~\ref{table:0}).
Finally, we obtained the Tully-Fisher distances using the calibration of \citet{Kourkchi2020} in the $r$-band, after correcting for the effects of reddening and inclination \citep[see][]{Kourkchi2019,Tully2013}. Results are shown in Table~\ref{table:0}. For D4 (LEDA1737674) we obtained a distance of only 16~Mpc, while the rest ranges from 36 to 53~Mpc, with an average of 46~Mpc and standard deviation of 7~Mpc, in good agreement with the Hubble flow value.

We do not deem the Tully-Fisher distance estimate of D4 to be realistic, because it would have too great a peculiar velocity at that distance. Rather, we think that the star forming regions for this small dwarf galaxy make an accurate inclination estimate unreliable. On the other hand, the individual distances of the other galaxies do not deviate more than 3-$\sigma$ from the Hubble flow value.

Placed at a fiducial distance of $42\pm7$~Mpc, NGC~2750 resembles a low-mass spiral galaxy. With a total magnitude of $M_V=-21$, it would have a stellar mass of $M_* \sim 2-3.5 \times 10^{10} M_\odot$ \citep[see also][]{Paudel2021}, assuming a mass-to-light ratio between 1 and 1.7 (the upper value being that of the Milky Way; \citealp{Bland-Hawthorn+Gerhard2016}). If we also assume a gas-to-stellar mass ratio of 0.5 (\citealp{Paudel2021}, but also \citealp{Baldry2008}), we would have a baryonic mass between $M_{\rm bary}=3-5 \times 10^{10} M_\odot$. Such values would typically correspond to a viral radius of 150~kpc \citep{Bullock+Boylan-Kolchin2017}. This translates into a projected radius of $\approx15$~arcmin, greater than the projected distance between the analysed galaxies and NGC~2750 itself. Therefore, by adding up this evidence, we can assume that the previously known galaxies around NGC~2750 are part of the same group and are likely to be its satellites as well.

\section{Low surface brightness features around NGC~2750}
\label{sec:analysis}

\begin{figure*}
    \centering
    \includegraphics[width=.45\textwidth]{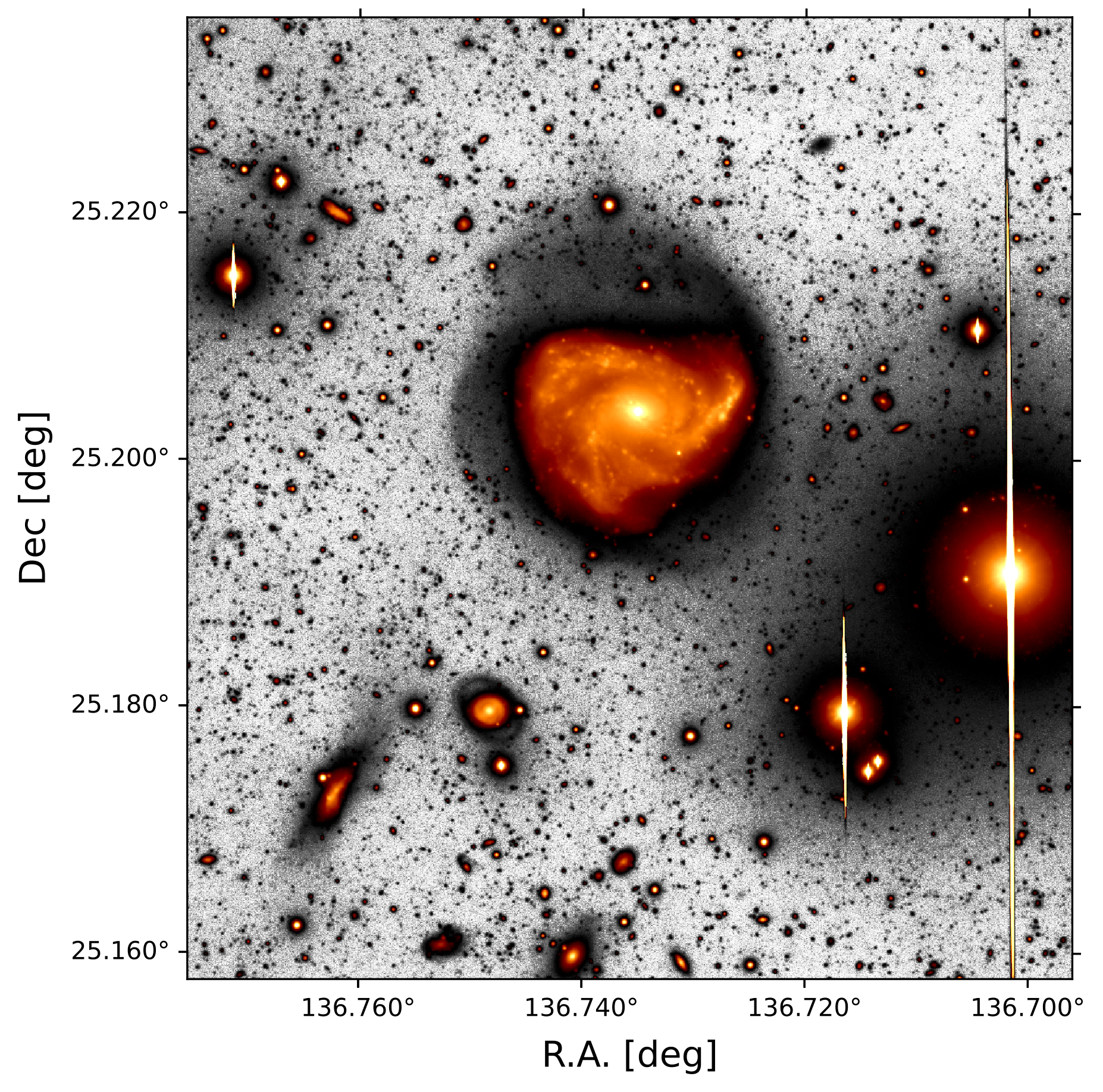}
    \includegraphics[width=.45\textwidth]{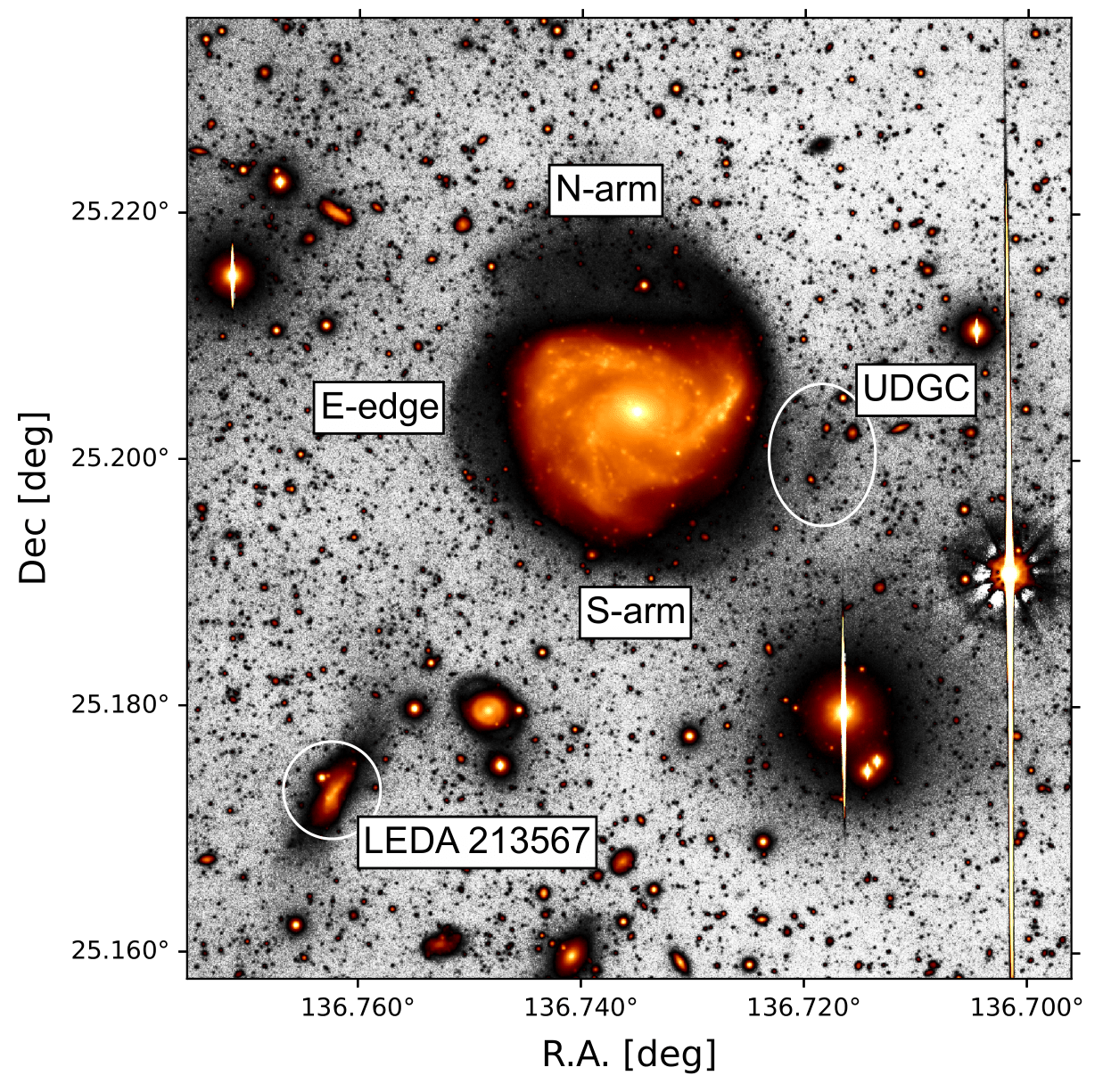}
    \caption{Scattered light removal. The final co-added image in the \textit{r}-band before (\textit{left}) and after (\textit{right}) subtracting the scattered light produced by the brightest star in the field. North is up and east to the left.}
    \label{fig:N2750_LSB}
\end{figure*}

The quality and depth of our LBT images allowed us to search for low surface brightness features around NGC~2750. 
These include the presence of tidal features around the central host and that of candidate satellites around the FoV. 

The outskirts of NGC~2750, however, are contaminated by the scattered light of a close-by bright foreground star, as shown in Fig.~\ref{fig:N2750_LSB}. 
This area was also of particular interest to us due to the presence of an overdensity visible to the west, which could be an independent satellite.
Therefore, the contribution of the foreground star had to be removed before further analysis of the central host and its surroundings could be carried out.

\subsection{Scattered light removal}
Several foreground stars are present in the FoV of our images, acting as sources of scattered light. The brightest star present (BD+25~2039 with $V\sim 11$) is also the closest one to NGC~2750. The scattered light produced by this star needed to be removed in order to correctly analyse the light profile of the central host and inspect the over-density found on its western side.

We accounted for the bright star by extracting its light profile with \texttt{photutils.profiles}, making circular apertures around its centroid. We then fitted the light profile with a power law of the form $\beta R^\alpha$ for both filters (with $\alpha$ and $\beta$ the constant parameters to be fitted), which was then converted into a 2D circular model that we subtracted from the original images \citep[see also][]{Golini2024}.

The scattered light removal process worked best for the \textit{r}-band, while for the \textit{g}-band it over-subtracts the central parts of the star due to the different saturation pattern (see Fig.~\ref{fig:scattered_light}). In both cases, though, the contribution of the scattered light is removed well, as shown in Fig.~\ref{fig:N2750_LSB} (right panel), leaving a background level of the order of the average sky in the image.

\subsection{Tidal features around NGC~2750}
The removal of the scattered light produced by BD+25~2039 allowed us to have a closer look at the outskirts of NGC~2750.
In particular, as shown in Fig.~\ref{fig:N2750_LSB}, we find extensive tidal perturbations coming off the north and south sides, together with a sharp-edged arm to the east, which could be a possible shell.
Such features might be caused by a close fly-by between the host and its satellite LEDA~213567 located in the south-east corner, that also shows signs of tidal tails in the direction of the central galaxy. However, we do not detect a connecting stream between the two galaxies.

The scattered light removal allowed us also to better highlight the over-density on the west side of NGC~2750. This feature appears on visual inspection as detached from the host. However, it is unclear whether this is an independent system, perhaps a satellite galaxy, or a tidal extension of the spiral arm of NGC~2750. A detailed analysis is presented in Sect.~\ref{sub-sec:udg}. 

Finally, the modelling of NGC~2750 described in Sect.~\ref{sec:distance} revealed no sign of a past merger event that could be related to the detected co-rotation signal. In particular, the analysis of its outer surface brightness profile showed no signs of a massive halo or tidal features until $\mu_r\sim29$~mag\,arcsec$^2$, after which we detected a sharp decline to background levels. The observed variations in its axis ratio and position angle are instead more likely due to the interaction with the aforementioned satellites. 
However, we cannot exclude the presence of traces of a past merger around NGC~2750 that have a sufficiently low surface brightness that our analysis cannot detect them.

\subsection{A possible ultra-diffuse galaxy next to NGC~2750?}
\label{sub-sec:udg}

\begin{figure*}
    \centering
    \includegraphics[width=\textwidth]{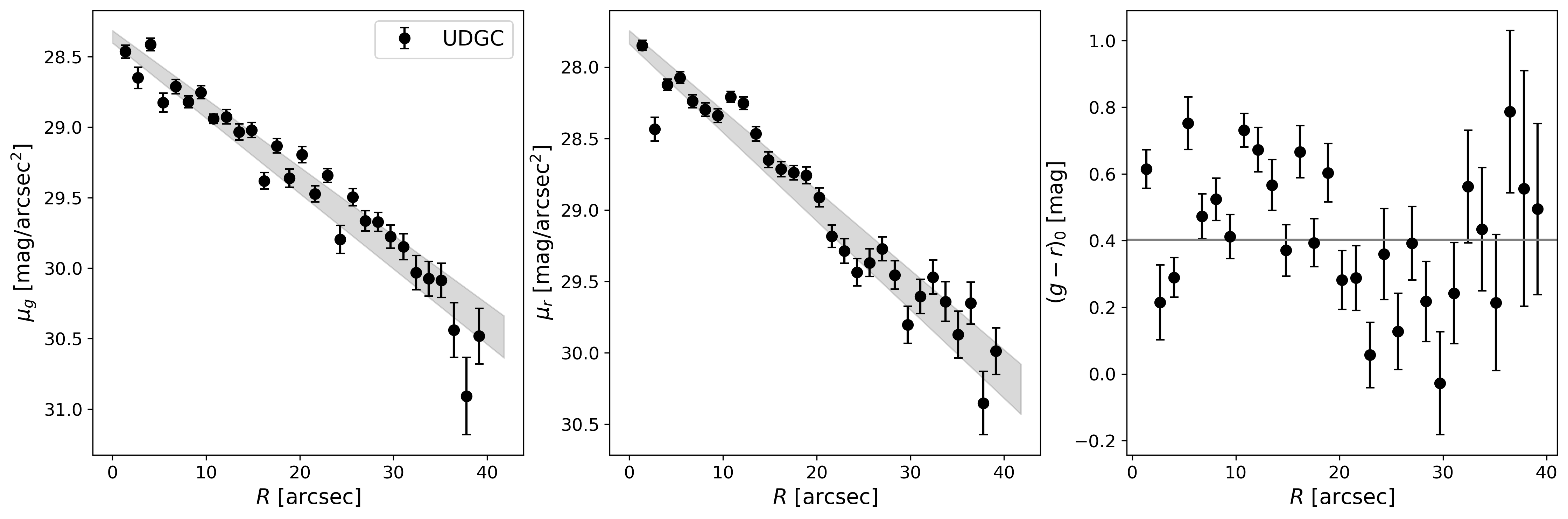}
    \caption{Radial surface brightness profile in the \textit{g}- and \textit{r}-bands (left and middle panels), together with the de-reddened colour profile (right panel) for the UDG candidate found near NGC~2750, where the horizontal solid line represents the median colour value.}
    \label{fig:SB_UDG}
\end{figure*}

We analyse here the over-density detected on the western side of NGC~2750, assuming initially that it is an independent satellite system.
To calculate its photometric properties, we re-binned the data by a factor of two (i.e. mapping $2\times2$ pixels onto 1 pixel using the average value) to increase the S/N. We masked the possible sources of contamination using the \textit{detect\_sources} function of \texttt{photutils.segmentation}, assuming a 3$\sigma$ threshold (see Fig.~\ref{fig:SB_UDG_Ellipse}). We then used the masked image to refine the local background estimation, subtracting the median value.

We continued by using the \textit{Ellipse} function of \texttt{photutils.isophote} to obtain the structural parameters and calculate the radial surface brightness profiles. We ran \textit{Ellipse} starting with an initial value for the centre, radius of the semi-major axis, ellipticity and position angle, allowing the code to find the most suitable parameters out to 35\arcsec, with a step of 1.35\arcsec (see Fig.~\ref{fig:SB_UDG_Ellipse}). 
The outermost radius was chosen to avoid contamination from the central host.

Finally, we modelled the extracted surface brightness profiles in both bands using the following Sérsic \citep{Sersic1968} equation:
\[
\mu(R) = \mu_0 + \frac{2.5}{\rm ln(10)}\left ( \frac{R}{h} \right )^{1/n}
\]
where $\mu_0$ is the central surface brightness, $h$ is the scale length and $n$ the Sérsic index \citep[see also][]{Graham+Driver2005}.
We corrected for the Galactic extinction, here and in the following, using the coefficients: $A_g = 0.12$ and $A_r = 0.08$ \citep{Schlafly+Finkbeiner2011}.

The light profiles shown in Fig.~\ref{fig:SB_UDG} (two leftmost panels) resulted well fitted by an exponential law (i.e. $n=1$), with an effective radius of $R_e\sim30$~arcsec, a central surface brightnesses of $\mu_0\sim28$~mag\,arcsec$^{-2}$ and an integrated magnitude of $m\sim19.5$~mag, in both bands. These values denotes already an extremely faint and extended system, compatible with being a satellite galaxy. A complete summary of the obtained photometric and structural values is reported in Tables~\ref{table:1} and \ref{table:2}. 

On the other hand, the radial colour profile in Fig.~\ref{fig:SB_UDG} (right panel), although noisy reveals a generally red system, which visually shows no sign of ongoing star formation. Since the integral of the brightness profile may be biased by the contamination from the central galaxy, we assume in the following as the colour index the median $(g-r)_0=0.4$, instead of the difference between the total magnitudes.

The \textit{Ellipse} analysis provided constant results with radius, with no evidence of structural asymmetry, torsion or other signs of tidal disruption. However, the high ellipticity of the system ($\epsilon=1-b/a=0.5$) and its proximity to the host suggest that it may be currently undergoing some degree of tidal disturbance. As discussed in Sect.~\ref{sec:discussion}, these evidences combined with the photometric properties of the system, make it more similar to a candidate ultra-diffuse galaxy (UDGC, as noted hereafter) than to a dwarf galaxy.

\begin{table*}
    \caption{Structural parameters of the dwarf galaxy candidates in the surveyed region around NGC~2750.}
    \label{table:1}
    \centering          
    \begin{tabular}{c c c c c c c c c c}    
        \hline\hline
        Name & R.A. & Dec. & CL & P.A.$_g$ & P.A.$_r$ & $\epsilon_g$ & $\epsilon_r$ & $n_g$ & $n_r$ \\ 
        & (deg) & (deg) &  & (deg) & (deg) &  &  &  & \\
        \hline  
        DGC0 & $136.41629$ & $+25.48070$ & $++$ & $115\pm2$  & $115\pm2$  & $0.37\pm0.02$ & $0.35\pm0.02$ & $0.64\pm0.02$ & $0.66\pm0.01$ \\ 
        DGC1 & $136.42515$ & $+25.55186$ & $+$  & $175\pm4$  & $175\pm6$  & $0.40\pm0.05$ & $0.40\pm0.06$ & $0.43\pm0.05$ & $0.44\pm0.09$ \\
        DGC2 & $136.55145$ & $+25.55524$ & $-$  & $165\pm5$  & $165\pm10$ & $0.50\pm0.07$ & $0.25\pm0.10$ & $0.75\pm0.09$ & $0.86\pm0.10$ \\
        DGC3 & $136.58624$ & $+25.29011$ & $-$  & $160\pm7$  & $160\pm10$ & $0.50\pm0.09$ & $0.50\pm0.11$ & $0.53\pm0.06$ & $0.77\pm0.10$ \\
        DGC4 & $136.57357$ & $+25.27167$ & $+$  & $5\pm14$   & $0\pm6$    & $0.30\pm0.14$ & $0.40\pm0.07$ & $0.78\pm0.11$ & $0.75\pm0.15$ \\
        DGC5 & $136.30185$ & $+25.24323$ & $-$  & $90\pm12$  & $90\pm43$  & $0.10\pm0.05$ & $0.05\pm0.07$ & $0.56\pm0.14$ & $1.3\pm0.2$   \\
        UDGC & $136.41700$ & $+25.42891$ & $++$ & $150\pm6$  & $150\pm4$  & $0.50\pm0.08$ & $0.50\pm0.05$ & $1.0\pm0.1$   & $1.0\pm0.1$   \\
        \hline                
    \end{tabular}  
    \tablefoot{The quantities listed in columns are: (1) candidate's name, (2-3) central coordinates, (4) detection significance, (5-6) position angle, (7-8) ellipticity (defined as $\epsilon=1-b/a$), and (9-10) Sersic index in \textit{g} and \textit{r}, with their respective uncertainties. In column (4), the signs $+$ and $-$ indicate our level of confidence in the visual detection of the candidate dwarf galaxies, with ($++$) a significant detection that appears symmetrical, diffuse and elliptical, ($+$) a similar detection but of lower surface brightness and ($-$) an uncertain detection due to a more ambiguous morphology.}
\end{table*}

\begin{table*}
    \caption{Photometric parameters of the dwarf galaxy candidates in the surveyed region around NGC~2750.}
    \label{table:2}
    \centering          
    \begin{tabular}{c c c c c c c c c c}    
        \hline\hline
        Name & $g_0$ & $r_0$ & $R_{e,g}$ & $R_{e, r}$ & $\mu_g(0)$ & $\mu_r(0)$ & $\langle \mu_{e,g}\rangle$ & $ \langle \mu_{e,r}\rangle$ \\ 
        & (mag) & (mag) & (\arcsec) & (\arcsec) & (mag/\arcsec$^2$) & (mag/\arcsec$^2$) & (mag/\arcsec$^2$) & (mag/\arcsec$^2$) \\
        \hline  
        DGC0 & $21.19\pm0.01$ & $20.69\pm0.01$ & $4.98\pm0.05$ & $4.86\pm0.02$ & $26.12\pm0.02$ & $25.54\pm0.01$ & $26.67\pm0.02$ & $26.12\pm0.01$ \\ 
        DGC1 & $22.56\pm0.05$ & $22.50\pm0.08$ & $5.51\pm0.20$ & $5.42\pm0.27$ & $27.99\pm0.04$ & $27.89\pm0.13$ & $28.25\pm0.04$ & $28.17\pm0.08$ \\
        DGC2 & $22.03\pm0.08$ & $21.81\pm0.07$ & $8.23\pm0.53$ & $7.34\pm0.49$ & $27.88\pm0.09$ & $27.23\pm0.07$ & $28.59\pm0.08$ & $28.12\pm0.10$ \\
        DGC3 & $22.57\pm0.06$ & $21.70\pm0.09$ & $4.51\pm0.19$ & $5.22\pm0.34$ & $27.44\pm0.04$ & $26.54\pm0.10$ & $27.83\pm0.05$ & $27.28\pm0.07$ \\
        DGC4 & $22.63\pm0.06$ & $22.10\pm0.10$ & $5.56\pm0.24$ & $6.32\pm0.53$ & $27.58\pm0.17$ & $27.37\pm0.18$ & $28.35\pm0.06$ & $28.08\pm0.12$ \\
        DGC5 & $23.57\pm0.09$ & $23.10\pm0.05$ & $2.97\pm0.16$ & $2.55\pm0.10$ & $27.49\pm0.31$ & $25.50\pm0.31$ & $27.93\pm0.14$ & $27.12\pm0.14$ \\
        UDGC & $19.71\pm0.07$ & $19.46\pm0.08$ & $35.9\pm1.78$ & $31.0\pm1.61$ & $28.36\pm0.04$ & $27.79\pm0.05$ & $29.48\pm0.04$ & $28.91\pm0.05$ \\
        \hline                
    \end{tabular}
    \tablefoot{The quantities listed in columns are: (1) candidate's name, (2-9) de-reddened total integrated magnitudes, effective radii, central and mean effective surface brightnesses in the \textit{g}- and \textit{r}-bands, with their respective uncertainties. Photometric values were corrected with the following Galactic extinction: $Ag = 0.117$~mag and $Ar = 0.081$~mag \citep{Schlafly+Finkbeiner2011}.}
\end{table*}

\subsection{New candidate satellites in the field of view}

We searched for new candidate satellites by visually inspecting the surveyed area around NGC~2750. To increase the S/N we performed a $4\times4$ rebinning of the data. Two of the authors (ST and OM) inspected the images independently, while the rest of the co-authors confirmed the results. The criteria followed were that candidate dwarf galaxies (DGCs) had to be detected in both bands, with their significance being established according to the confidence scale given in Table~\ref{table:1}.

We confirmed the DGC reported by \citet[][their D7 noted as DGC~0 throughout the text]{Paudel2021}, and found five additional candidates. 
The DGC cutouts are shown in Fig.~\ref{fig:lbc_DG}. 
Due to their very low surface brightness, for three of them (noted as DGC~2, 3 and 5) we report an initial uncertain classification (see Table~\ref{table:1}). 
We have also checked that there are no known dwarf galaxies, or possible background contaminants, around the coordinates of the discovered systems.

In order to calculate the photometric properties of the DGCs, we followed the same procedure as for the UDGC, reported in the previous section. In this case, we have used the curve-of-growth method to find the outermost radius up to where to run \textit{Ellipse}. This radius is in fact defined as the point at which the cumulative magnitude, as we move away from the inspected galaxy, reaches a plateau (i.e. hit the background level). 
The extracted surface brightness and colour profiles of the DGCs are shown in Fig.~\ref{fig:SBrad}, while their integrated properties are summarised in Tables~\ref{table:1} and \ref{table:2}. 
These systems are characterised by a very low surface brightness and a generally red colour profile. In particular, they have effective radii that extend no further than $R_e\sim5\arcsec$, comparable mean effective surface brightness of $\langle \mu_e\rangle\sim27.5$~mag\,arcsec$^{-2}$ and Sersic indexes of $n\sim0.6$. 

With regard to our detections, we carried out a qualitative comparison with the LIGHTS survey \citep{Trujillo2021,Zaritsky2024}, with which we share a similar observational strategy. We found broad agreement in terms of photometric properties, although our candidates are on average fainter and smaller than those examined by \citet{Zaritsky2024}. In particular, DGC~5 and UDGC stand out for being at the two extremes of compactness and diffusion, respectively. We note, however, that our observations are deeper than those conducted in \citet{Zaritsky2024} and comparable to those of the pilot study presented in \citet{Trujillo2021}. In fact, our DGCs resemble in photometric terms the faintest system they identified (NGC~1042-LBT1).

In the discussion section, we determine whether the identified candidates, once placed at the distance of NGC~2750, have physical properties compatible with those of confirmed dwarf galaxies in the local Universe. We should emphasise, however, that even this comparison is provisional and subject to future spectroscopic confirmation.

\section{Discussion}
\label{sec:discussion}

We revealed a number of low surface brightness features around NGC~2750, among which the discovery of six new satellite candidates. In the following section we compare them with other known dwarf galaxies in the local Universe. 
We also discuss their impact on the luminosity function of NGC~2750 and the phase-space correlation. 
In the following, we assume that NGC~2750 is at a fiducial distance of 42~Mpc (corresponding to the Hubble flow), to which we associate an uncertainty of 7~Mpc (so as to account for the scatter of individual distance measurements; see Sect.~\ref{sec:distance}).

\subsection{Properties of the dwarf galaxy candidates}

\begin{figure*}
    \centering
    \includegraphics[width=\textwidth]{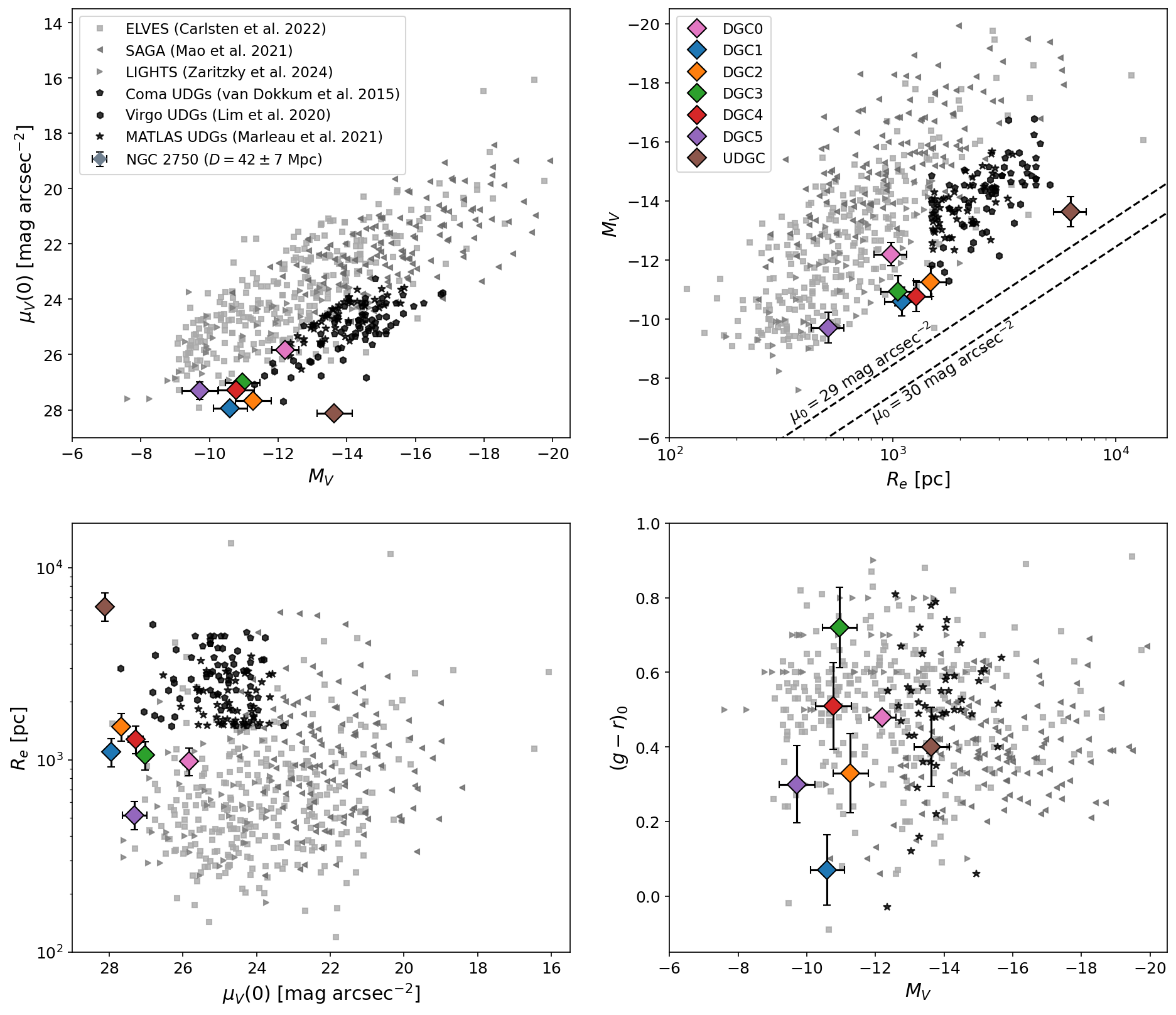}
    \caption{Scaling relations for the candidate satellites of NGC~2750 compared to the distribution of values for known dwarf and ultra-diffuse galaxies. In the \textit{upper left panel} is shown the absolute visual magnitude $M_V$ vs the central surface brightness $\mu_V(0)$ (in units of mag\,arcsec$^{-2}$); in the \textit{upper right panel}, $M_V$ vs the effective radius $R_e$ (in units of parsec), with overlaid lines of constant $\mu_V(0)$ (at 29 and 30 mag\,arcsec$^{-2}$); in the \textit{lower left panel}, $\mu_V(0)$ vs $R_e$; in the \textit{lower right panel}, $M_V$ vs the de-reddened colour index $(g-r)_0$. The candidate satellites of NGC~2750 are marked as coloured diamonds, assuming a group's distance of $42\pm7$\,Mpc. They are compared to the dwarf galaxies from the ELVES \citep[][dark-gray squares]{Carlsten2022}, SAGA \citep[][leftward dim-gray triangles]{Mao2021}, and LIGHTS \citep[][rightward gray triangles]{Zaritsky2024} surveys, together with the ultra diffuse galaxies of the Coma \citep[][black diamonds]{vanDokkum2015} and the Virgo \citep[][black hexagons]{Lim2020} clusters and in the MATLAS survey \citep[][black stars]{Marleau2021}.}
    \label{fig:scaling-relations}
\end{figure*}

A standard approach to establish the group membership of new candidate satellites is to make a comparison with known systems on a diagram showing a particular scaling relation \citep{Jerjen2000,Chiboucas2009,Muller2017,Muller2018b,Habas2020}. 
In the case of NGC~2750, this allows us not only to distinguish satellite galaxies from background contamination, but also to validate the goodness of the group distance estimation (see Sect.~\ref{sec:distance}).

We performed a comparison with the compilation of confirmed dwarf galaxies from the ELVES \citep{Carlsten2022}, SAGA \citep{Mao2021}, and LIGHTS \citep{Zaritsky2024} surveys, together with the UDGs found in the Coma and Virgo clusters \citep{vanDokkum2015,Lim2020} and in the MATLAS survey \citep{Marleau2021}.\footnote{All photometric values expressed in the \textit{ugriz} system were transformed into the Johnson system following the conversion formulas of \citet{Jordi2006}. We note that for the UDGs case, \citet{vanDokkum2015} and \citet{Lim2020} only provided magnitudes in the \textit{g}-band. Therefore, to transform them into \textit{V}-band magnitudes, we assumed a colour index of $(g-r)_0=0.4$~mag, typical of these systems \citep[see][]{Marleau2021}.}

The considered compilations cover a wide range of distances (from the Local Group out to 40~Mpc) and host's properties. The comparison with the LIGHTS survey is particularly useful, considering that we share the same instrument and observational set-up.
The range of distances covered by the survey is $10<D[{\rm Mpc}]<30$, with a median value of $\sim15$~Mpc. 
We note that the survey has a detection limit of $M_V\sim-7.5$, while SAGA and ELVES have $M_V\sim-11.5$ and $-9.0$, respectively. 

The comparison is shown in Fig.~\ref{fig:scaling-relations}, where we highlight the relations between the absolute visual magnitude $M_V$ and the central surface brightness $\mu_V(0)$ (upper left panel), between $M_V$ and the effective radius $R_e$ (upper right), and between $\mu_V(0)$ and $R_e$ (lower left), together with the de-reddened colour index $(g-r)_0$ as a function of $M_V$ (lower right). 

In all panels, the DGCs follow the distribution of known systems, which supports their bona-fide classification as dwarf galaxies. Furthermore, considering that NGC~2750 is located in a low-density environment with a low probability of contamination from other galaxy groups \citep[see the discussion in][]{Paudel2021}, it is likely that the DGCs could also be satellites of NGC~2750. We defer to future spectroscopic observations to confirm this aspect.

\subsection{Properties of the ultra diffuse galaxy candidate}
From Fig.~\ref{fig:scaling-relations}, we see that the UDGC lies at the faint end of the distribution of UDG values. Typically, UDGs are defined to have $R_e>1.5$~kpc, $\mu_V(0)>24$~mag\,arcsec$^{-2}$ and $(g-r)_0\sim0.5$~mag \citep{vanDokkum2015,Lim2020,Marleau2021}, making the UDGC compatible with this class of systems. However, the UDGC is extremely faint with $\mu_V(0)\sim28$~mag\,arcsec$^{-2}$, which would make it one of the faintest galaxies in its category, if confirmed to be so. Its properties are more similar to those of Antlia~II \citep{Torrealba2019,Ji2021}, a satellite dwarf galaxy of the Milky Way with $R_e\sim2.5$~kpc and $\mu_V(0)\sim29$~mag\,arcsec$^{-2}$, which is probably in the process of being tidally disrupted. The similar ellipticity ($\epsilon\sim0.5$) to Antlia~II and proximity to its host galaxy may indicate that the UDGC is on an infall orbit towards its host. 

We note that the UDGC case is also analogous to that of two UDGs reported by \citet{Bennet2018}. These systems are dominated by an old stellar population, reside in a group environment and seem to be associated with tidal material originating from the interaction with a larger galaxy. It has been suggested that UDGs that share these properties may have formed through interactions between galaxies, which stripped them of their gas and caused their stellar population to expand \citep[e.g.][]{Carleton2019,Tremmel2020}. They could also be tidal dwarf galaxies \citep[e.g.][]{Duc2012}, born from the recycling of gas-rich material originating from the debris of a merger/interaction between galaxies. In the UDGC case, a high spatial resolution follow-up would allow us to identify its population of globular clusters and to discern between different formation scenarios \citep[e.g.][]{Saifollahi2022,Fielder2024}.

\subsection{The luminosity function of the NGC~2750 system}

\begin{figure*}
    \centering
    \includegraphics[width=.48\textwidth]{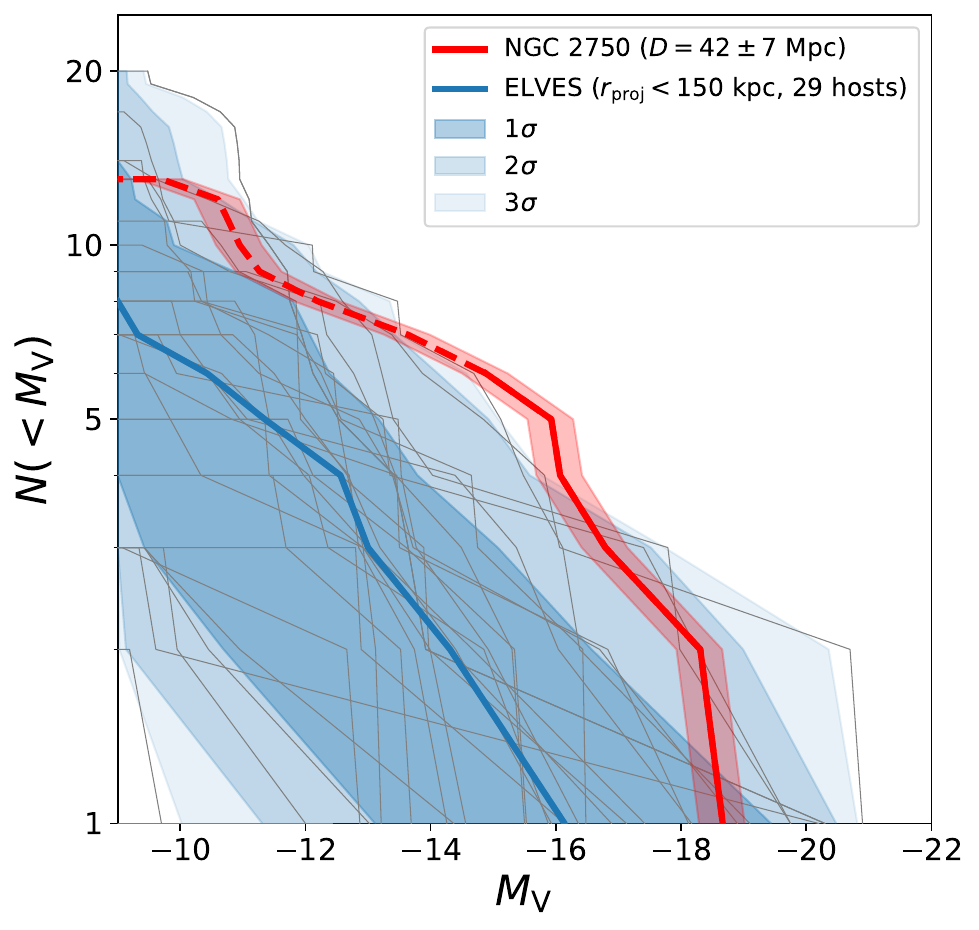}
    \includegraphics[width=.50\textwidth]{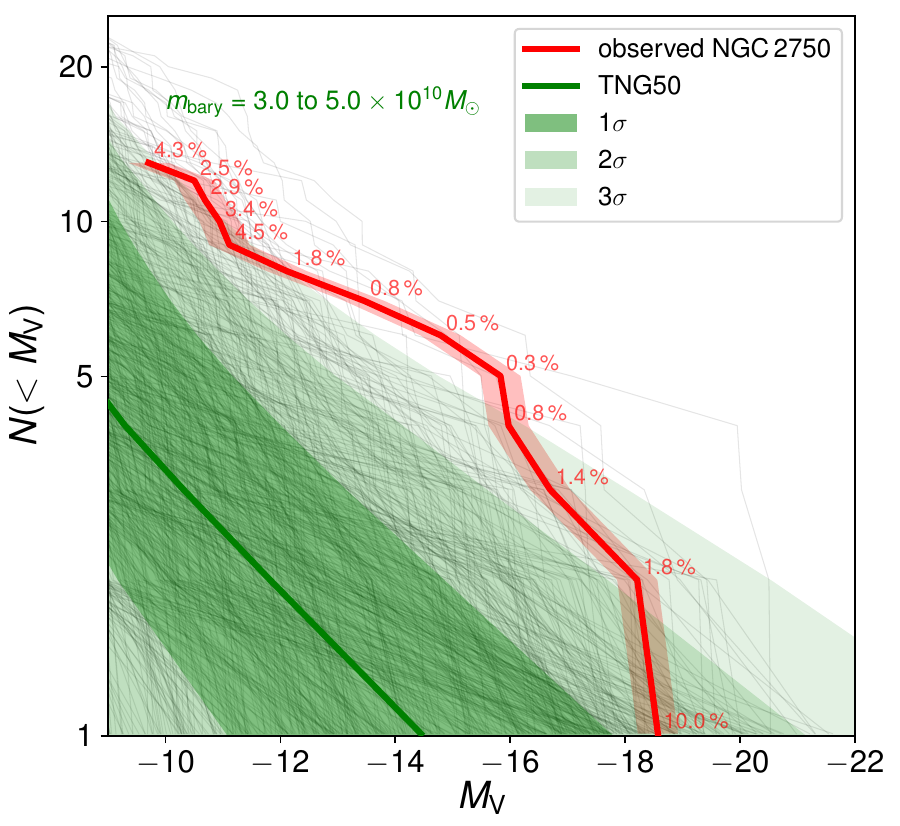} 
    \caption{Luminosity function of the NGC~2750 system compared with those of observed and simulated systems. \textit{Left:} comparison with the observed systems studied by the ELVES survey (gray lines). Values for the NGC~2750 system are marked with a red line, with the dashed part indicating the values of the DGCs, while the shaded area accounts for the distance uncertainty. The blue line represents the median distribution of the ELVES survey, with the shaded areas being the 1, 2, and 3$\sigma$ confidence intervals. The lower $M_V$ limit coincides with the ELVES survey completeness limit of $-9$~mag \citep{Carlsten2022}. \textit{Right:} comparison with the simulated systems selected from TNG50 (gray lines). Again, the green line represents the median distribution, with the shaded areas being the 1, 2, and 3$\sigma$ intervals.}
    \label{fig:lum-function}
\end{figure*}

We can now examine the luminosity function of the NGC~2750 system, comparing it with those of other observed and simulated systems, as shown in Fig.~\ref{fig:lum-function}. In the left panel of the figure, we show a comparison with the satellite distributions of the galaxy systems analysed by the ELVES survey \citep{Carlsten2022}. The distribution of NGC~2750 is shown taking into account the uncertain association of the DGCs and the uncertainty on the distance measurement.

The ELVES systems have a complete detection of satellites with $M_V<-9$ within a projected distance of the central host of $r_{\rm proj}=150$~kpc. This radius corresponds to the assumed virial radius of NGC~2750 and is smaller than the maximum projected distance in our FoV. This last point is particularly important because one effect of the distance uncertainty is to vary the FoV in physical units.

From the figure it can be seen that the distribution of the NGC~2750 system shows more luminous members (i.e. those with a confirmed spectroscopic association) than the majority (i.e. $\gtrsim3\sigma$) of the survey distributions, regardless of the inclusion of the DGCs (marked with a dashed line). In other words, although some of the hosts in ELVES have satellites that could be more luminous than those around NGC~2750, only one of these hosts (i.e. NGC~1023) have the same amount of bright satellites (i.e. $\ge6$, placing a limit at $M_V=-14.9\pm0.4$).

We note that a similar result is also obtained by comparison with the LIGHTS and SAGA surveys, although they do not have the same level of completeness within the FoV as ELVES \citep[see][]{Mao2021,Zaritsky2024}. However, this is no less striking as these surveys target hosts covering a wide range of stellar masses, from MW analogues (SAGA) to fainter systems down to LMC luminosities (LIGHTS and ELVES). It therefore appears that the NGC~2750 system has an excess of galaxies in its distribution at the high-luminosity range, where completeness biases are minimised.
 
Recently, \citet{Muller2024} and \citet{Kanehisa2024} reported, respectively, that the luminosity distribution of the M~83 group and the low-to-moderate density fields of MATLAS show an excess of dwarf galaxies compared to expectations from the IllustrisTNG hydrodynamic cosmological simulation. In particular, \citet{Muller2024} showed that this "too-many-satellites" problem is significant at the faint end of the luminosity distribution of M~83, whereas in our case the discrepancy with other observations occurs at the bright end. 

Similarly to \citep{Muller2024}, we performed a comparison with IllustrisTNG. In particular, we used the publicly available catalogue of TNG50-1 simulated galaxies at $z=0$ \citep{Nelson2019}. We selected primary halos analogous to NGC~2750 with baryonic masses between $M_{\rm bary}=3.0-5.0\times10^{10}M_\odot$ (see Sect.~\ref{sec:distance}). We also required that there be no other galaxies more massive than 50\% of the host within 300~kpc, in order to better resemble the environment around NGC~2750. We obtained 650 isolated analogues, which we mock observed with a random orientation at NGC~2750 distance. For each host, we selected sub-halos that are within a depth along the line of sight of 500~kpc and projected to lie within $0.275^\circ$\ around the host's centre (excluding the inner 10\% of the radius).

The results of the comparison with TNG50 is shown in the right panel of Fig.~\ref{fig:lum-function}. The great majority of simulated analogues has a luminosity distribution below the one observed for NGC~2750. In particular, the agreement is at most within $2\sigma$ only at the very bright and very faint ends. As well, we detect an excess of observed galaxies around $M_V\sim-15$ ($>3\sigma$), similar to the comparison with observed systems from ELVES. Indeed, only 0.5\% of simulated analogues shows more satellites at this limit. Therefore, the NGC~2750 system seems in fact unusually rich of bright satellites.

\subsection{The group coherent motion in the light of new satellite candidates}

\begin{figure*}
    \centering
    \includegraphics[width=.49\textwidth]{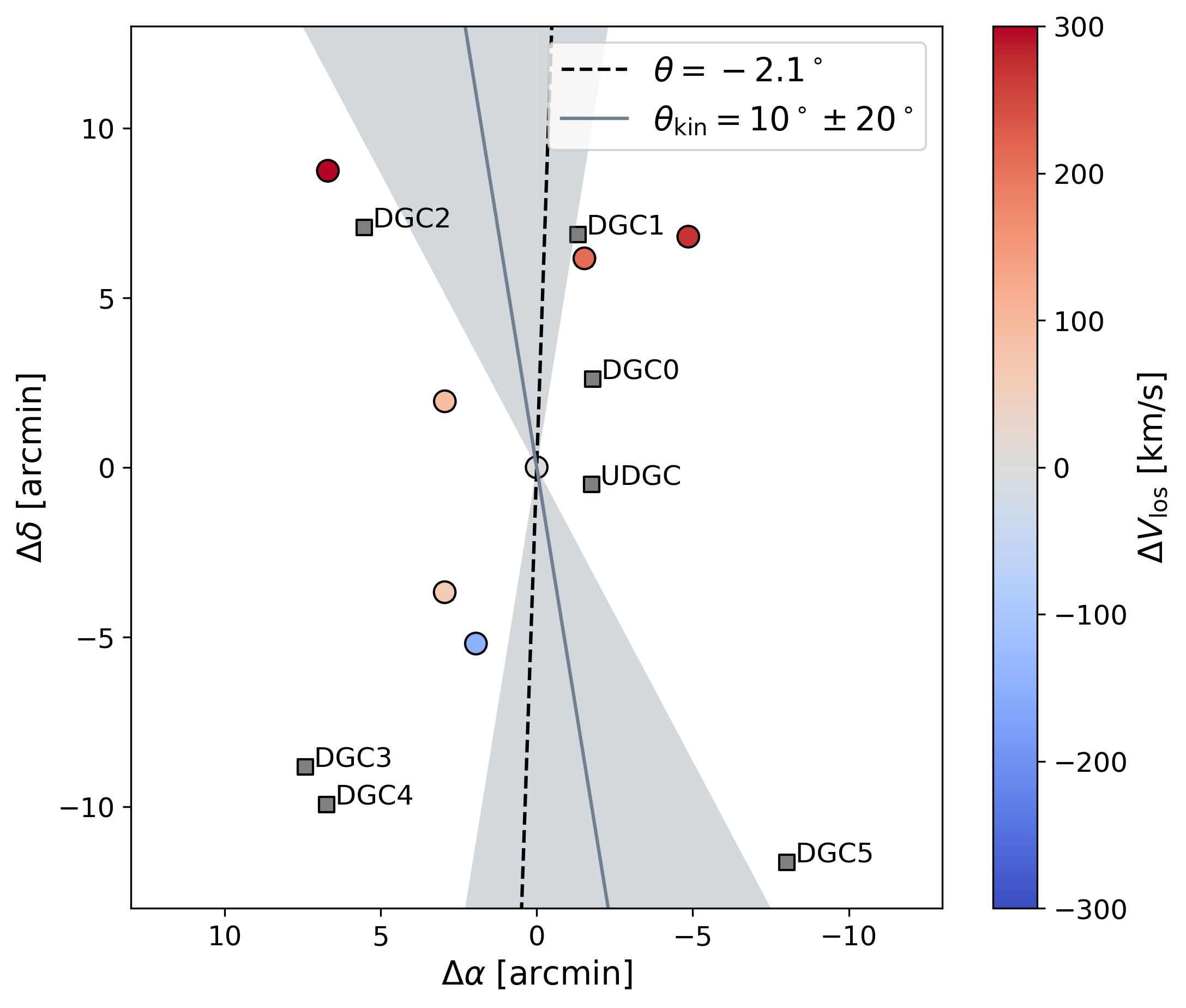}
    \includegraphics[width=.49\textwidth]{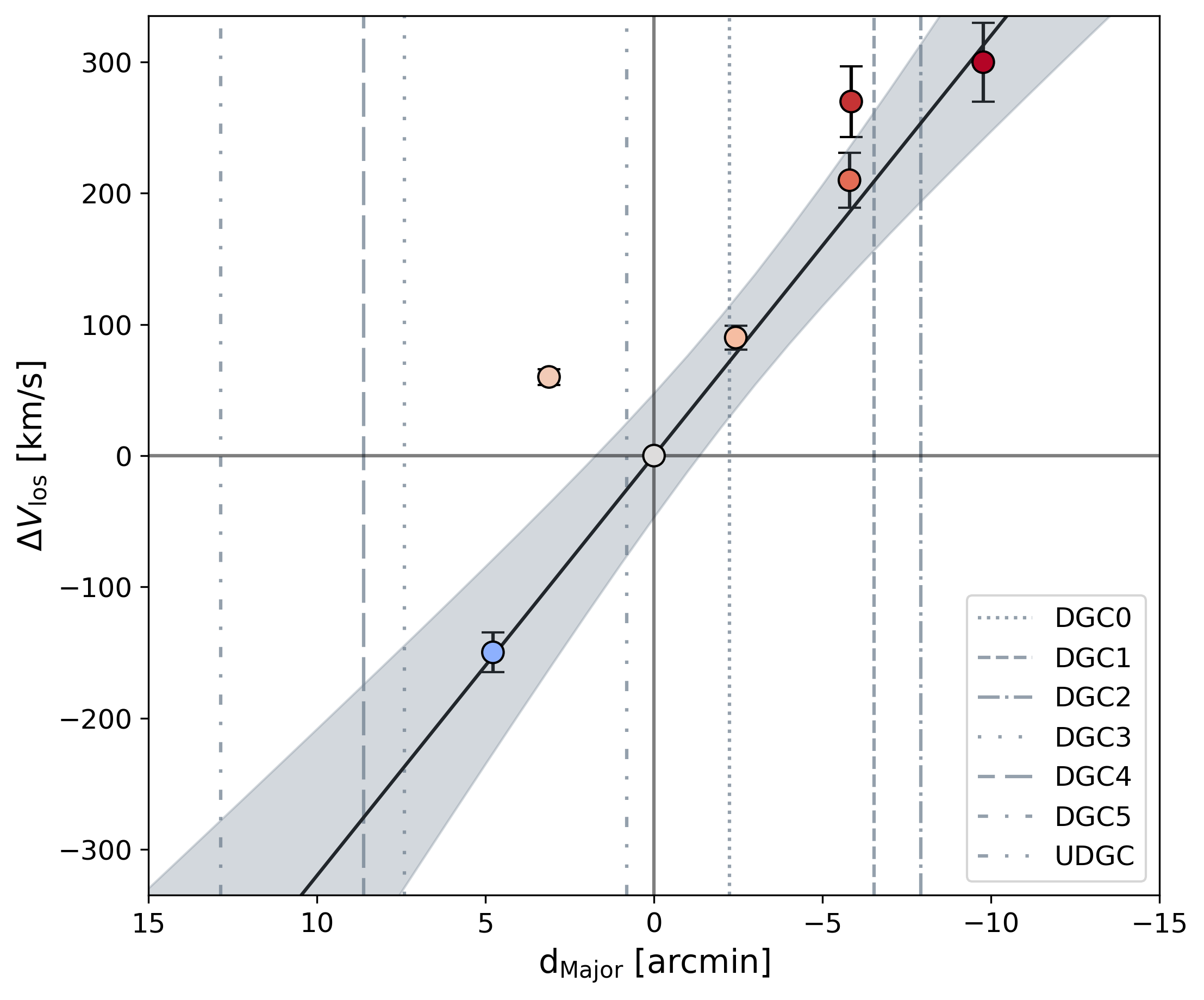}
    \caption{Phase-space diagrams of the NGC~2750 system. \textit{Left:} Spatial distribution of known members (circles) and candidates (squares) coloured (when measurements are available) according to their difference in line-of-sight velocity from that of the central host. \textit{Right:} line-of-sight velocity difference as a function of the on-sky distance along the major axis ($d_{\rm Major}$); the solid black lines is the best fit of the rotational model with the $1\sigma$ confidence interval, while the discontinuous vertical lines represent the $d_{\rm Major}$ values of the DGCs.}
    \label{fig:coherent-motion}
\end{figure*}

In their analysis of the NGC~2750 system, \citet{Paudel2021} reported that it shows a significant kinematic correlation between the group members. This led them to conclude that this system could be a low-mass version of the Centaurus~A group, which shows a well-established phase-space correlation \citep[e.g.][]{Muller2021}. We examine here whether this statement is true in light of the new candidate systems reported in this paper. We emphasise again that this analysis is provisional with respect to their future confirmation as satellites of NGC~2750.

Since one of the key aspects of a phase-space correlated system is its spatial flattening, we first measure how the NGC~2750 system changes with the addition of the DGCs assumed as its satellites. Performing a linear orthogonal distance regression fit, we measured a minor-to-major-axis root-mean-square (rms) flattening of the projected distribution of $b/a=0.67\pm0.12$, where the error represents the measured scatter depending on the inclusion of candidate satellites by applying a jackknife resampling. This would imply an inclination $i={\rm arccos}(b/a)=48^\circ\pm9^\circ$, if we were to assume that the underlying structure is indeed planar. The position angle of the major axis of the projected distribution (measured from north to east) is $\theta=-2.1^\circ$, as also shown in Fig.~\ref{fig:coherent-motion} (left panel). 

For comparison, considering only the previous sample studied by \citet{Paudel2021} we would have comparable results with $b/a=0.62\pm0.34$ and $\theta=1.9^\circ$. Therefore, the inclusion of the DGCs does not change significantly the spatial flattening of the system, which remains low in absolute values compared to that of known flattened distributions ($b/a \leq 0.5$, \citealp[e.g.][]{Pawlowski2021}).

Before inspecting how the kinematic correlation may be affected by the DGCs, we performed a Bayesian analysis to calculate the kinematic parameters of the NGC~2750 system. We followed the method reported in \citet{Taibi2018}, assuming a linear kinematic model with no intrinsic dispersion. 
Specifically, we assumed: $V_{\rm rot}=kR_i\,{\rm cos}(\theta-\theta_i)$, where $(R_i,\theta_i)$ are the radial distance from the central host and position angle (measured from north to east) of the i-galaxy satellite, 
$\theta$ is the position angle of the kinematic major axis, $V_{\rm rot}$
is the observed rotational velocity defined as the difference between the l.o.s.~velocities of the i-satellite and the host. 
The parameters to be fitted are the kinematic gradient $k$ and the position angle of the kinematic major axis $\theta$, respectively with the following priors ranges: $-100 < k [ {\rm km\,s^{-1}\,arcmin^{-1}}]< 100$ and $0\le \theta_k [^\circ]\le 180$. 

Results for the kinematic modelling were: $k=32\pm8$~km\,s$^{-1}$\,arcmin$^{-1}$ and $\theta_k=10^\circ\pm20^\circ$. In particular, we see that the position angle of the kinematic major axis is in agreement within the errors with that of the major axis of the projected distribution (see again the left panel of Fig.~\ref{fig:coherent-motion}). 

With regard to the observed kinematic correlation, in their analysis \citet{Paudel2021} performed a Pearson correlation test and found a coefficient $c=0.90$ and a p-value $=0.005$. However, with the DGCs inclusion it is clear that even a few of them, if confirmed as satellites, could enhance or diminish the significance of the correlation depending whether their line-of-sight velocities turn out to follow the trend of the previously known satellites or not. 

From Fig.~\ref{fig:coherent-motion}, we see that the DGCs are distributed close to the kinematic major axis (left panel) and divide evenly on the two sides of the derived position-velocity diagram (right panel). 
If we were to include only the two DGCs closest to NGC~2750 (i.e. DGC~0 and UDGC) and assign them random velocities extracted $10^4$ times from a uniform distribution with limits $[-300;300]$~km\,s$^{-1}$, we would still obtain a strong correlation with a p-value$<0.05$.
On the other hand, the inclusion of only one of the remaining DGCs with a velocity contrary to the present trend would decrease the correlation and lead the p-value to be $>0.05$. This is because these systems are located far from the central host and thus can have a greater impact on the velocity curve. 

To make any progress on understanding the dynamics of the NGC~2750 system and its kinematic coherence, it will be vital to obtain spectroscopic follow-ups. However, the most promising system in this case would be DGC~0 thanks to its brightness, which would make it an ideal target for an eight-meter class telescope, such as the LBT or the Very Large Telescope (VLT). All the other DGCs, although more discriminating for the kinematic correlation, have too low a surface brightness for the current multi-object and integral-field-unit facilities.

\section{Conclusions}
\label{sec:end}

We acquired deep photometric observations of the region surrounding NGC~2750 using the Large Binocular Telescope (LBT) to investigate the presence of a group of co-rotating satellite galaxies. Our results extend the known satellite system of this galaxy, while providing new information on the low surface brightness features surrounding NGC~2750.

Our analysis revealed tidal features extending from the arms of NGC~2750, indicating past interactions with its close satellites. However, the absence of a massive halo around the central host makes it unlikely that the observed co-rotation of the system is the product of a past major merger. 

The presence of the tidal features around NGC~2750 prevented us from improving the measurement of the galaxy's inclination, and thus its distance using the Tully-Fisher relation. Instead we modelled its satellite galaxies with available HI measurements, finding a group distance averaging around 40~Mpc, compatible with that obtained from the Hubble flow. 

We have identified six new candidate dwarf galaxies around NGC~2750, in addition to the already known satellites. These systems exhibit low surface brightness and generally red colour profiles. Their properties closely align with those of known dwarf galaxies in the local Universe, independently of the distance uncertainties. One candidate in particular exhibits features consistent with an ultra-diffuse galaxy, but could also be a dwarf galaxy undergoing tidal destruction.

The spatial distribution of the candidate satellites does not change (and thus supports) the substantial flattening of the system, which remains low compared in particular to those observed around the Milky Way or M31. 
The system exhibits a kinematic correlation similar to that of the Cen~A group, but the inclusion of the new candidates could cast doubt on the universality of this correlation, which can easily vanish once spectroscopic measurements are available for our additional candidate satellite galaxies.

Finally, the luminosity function of the NGC~2750 system indicates an excess of bright satellites compared to other galaxy groups studied in surveys such as ELVES, SAGA and LIGHTS.
We recall that a tension has recently arisen in the literature between observational data showing ‘too many satellites’ when compared with the predictions of cosmological simulations \citep{Muller2024,Kanehisa2024}. 
The NGC~2750 system could be another observational example of this problem.

\begin{acknowledgements}
We wish to thank the anonymous referee for the constructive feedback.
We thank G.~Golini, I.~Trujillo, and S.~Paudel for useful discussions and comments at various stage of this project.
MSP acknowledges funding via a Leibniz-Junior Research Group (project number J94/2020). 
OM is grateful to the Swiss National Science Foundation for financial support under the grant number PZ00P2\_202104.
MB, MJ, AL and SS acknowledge support by the Ministry of Science, Technological Development and Innovation of the Republic of Serbia under contract no. 451-03-136/2025-03/200002 with the Astronomical Observatory of Belgrade.
This research has made use of NASA’s Astrophysics Data System, VizieR catalogue access tool (CDS, Strasbourg, France, DOI: 10.26093/cds/vizier), and extensive use of python3.8 \citep{Python3}, including ipython \citep[v8.12,][]{ipython}, numpy \citep[v1.24,][]{NumPy-Array}, scipy \citep[v1.10,][]{SciPy-NMeth}, matplotlib \citep[v3.7,][]{Matplotlib}, astropy \citep[v5.2,][]{Astropy}, photutils \citep[v1.8][]{photutils}, and ccdproc \citep[v2.4][]{ccdproc}.
\end{acknowledgements}

%
%

\bibliographystyle{bibtex/aa} 
\bibliography{bibtex/biblio.bib} 


\begin{appendix}
\onecolumn

\section{Supplementary material}
\label{sec:apx1}

\begin{figure*}[h!]
    \centering
    \includegraphics[width=\textwidth]{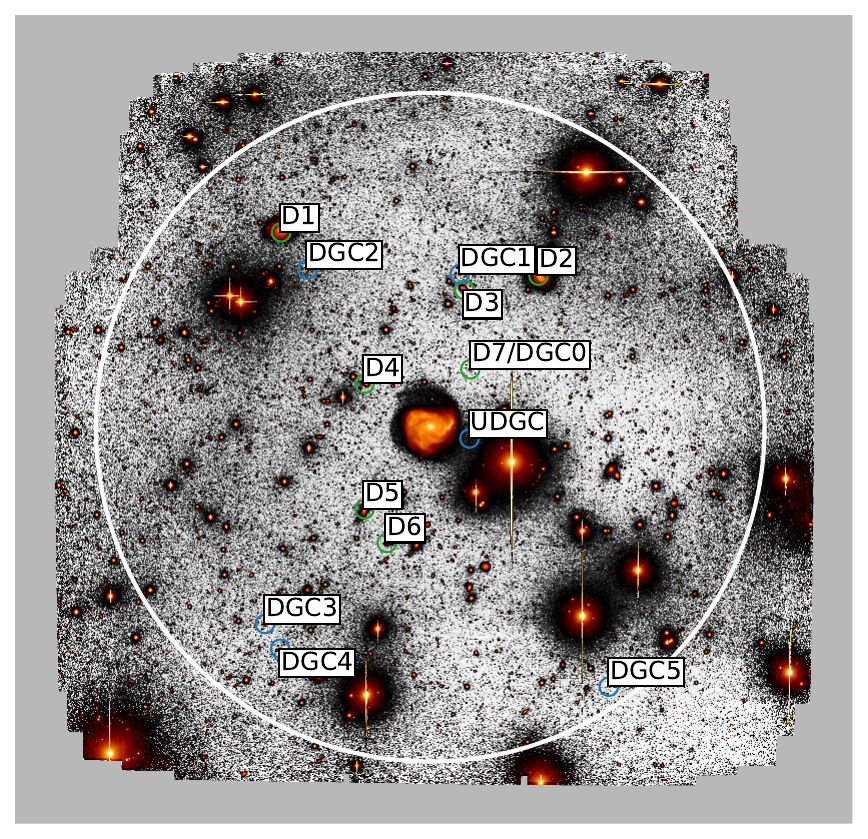}
    \caption{Co-added $(g+r)/2$ image of the surveyed area around NGC~2750. The position of the known and candidate galaxies is marked with the names given to them by \citet[][see their Table~1]{Paudel2021} and us (our Table~\ref{table:1}), respectively.
    The large white circle marks the host's virial radius (i.e. 150~kpc $\approx15$\arcmin), while the image FoV is $\approx35\arcmin\times30\arcmin$. 
    North is up and east to the left.}
    \label{fig:lbc_sats}
\end{figure*}

Here we report additional figures in support of the analysis reported in the main text. 
In particular, in Fig.~\ref{fig:lbc_sats} is shown the co-added $(g+r)/2$ image of the LBT surveyed area, with marked the names of known and candidate satellite galaxies around NGC~2750.

\begin{figure*}[h!]
    \centering
    \includegraphics[width=.75\textwidth]{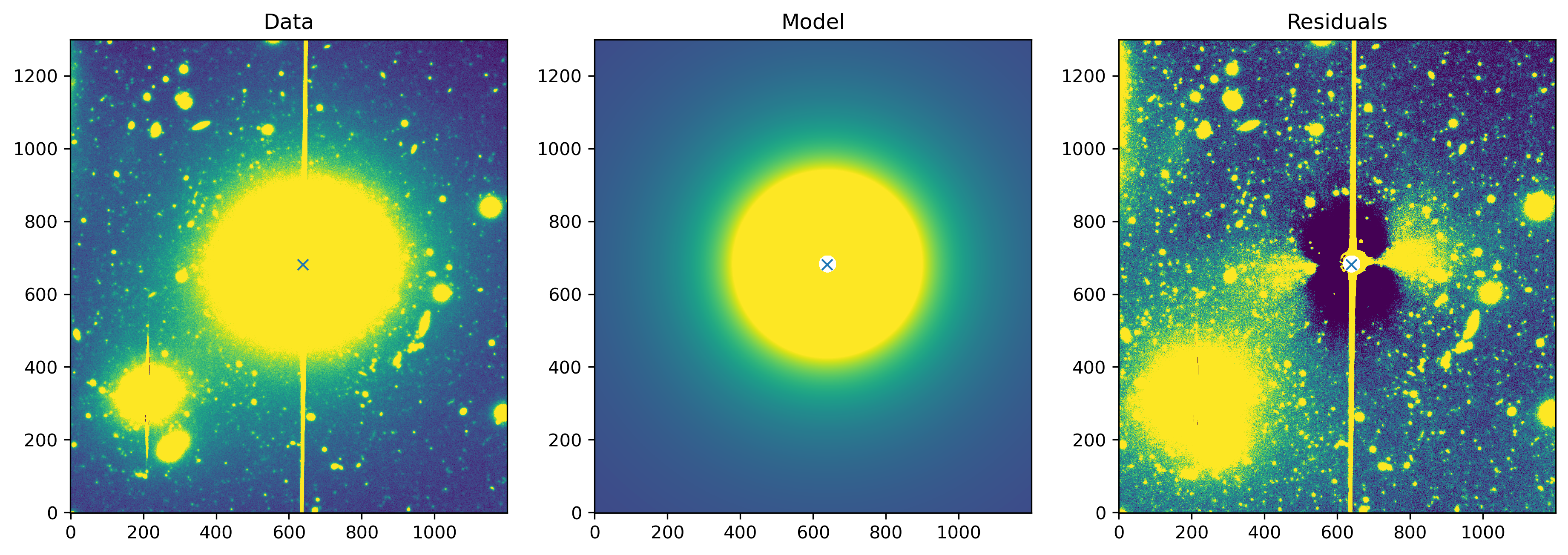}
    \includegraphics[width=.75\textwidth]{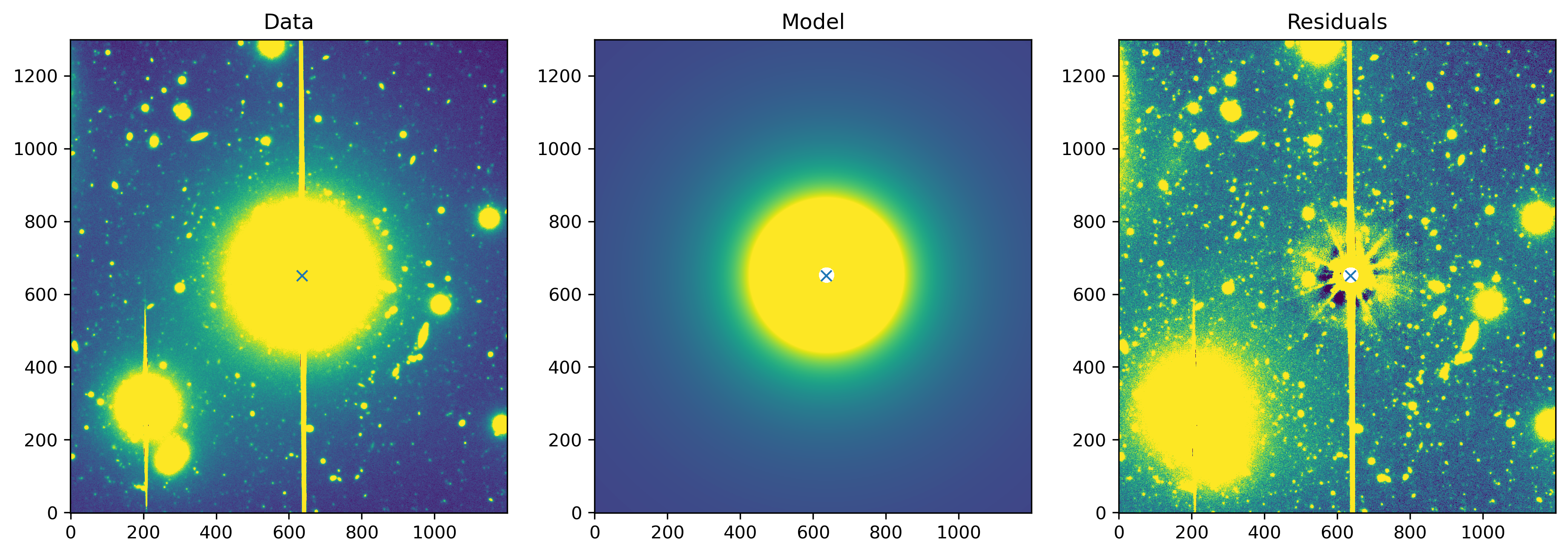}
    \caption{Scattered light removal. Cutouts of the area around the bright star BD+25~2039 as it appear in the data (\textit{left}), its modelling (\textit{centre}), and the residuals once having subtracted the model (\textit{right}), for the \textit{g}- (\textit{top}) and the \textit{r}- (\textit{bottom}) bands. Each image has a size of $\sim4.5\arcmin\times4.5\arcmin$. North is up and east to the left.}
    \label{fig:scattered_light}
\end{figure*}

\begin{figure*}[h!]
    \centering
    \includegraphics[width=.75\textwidth]{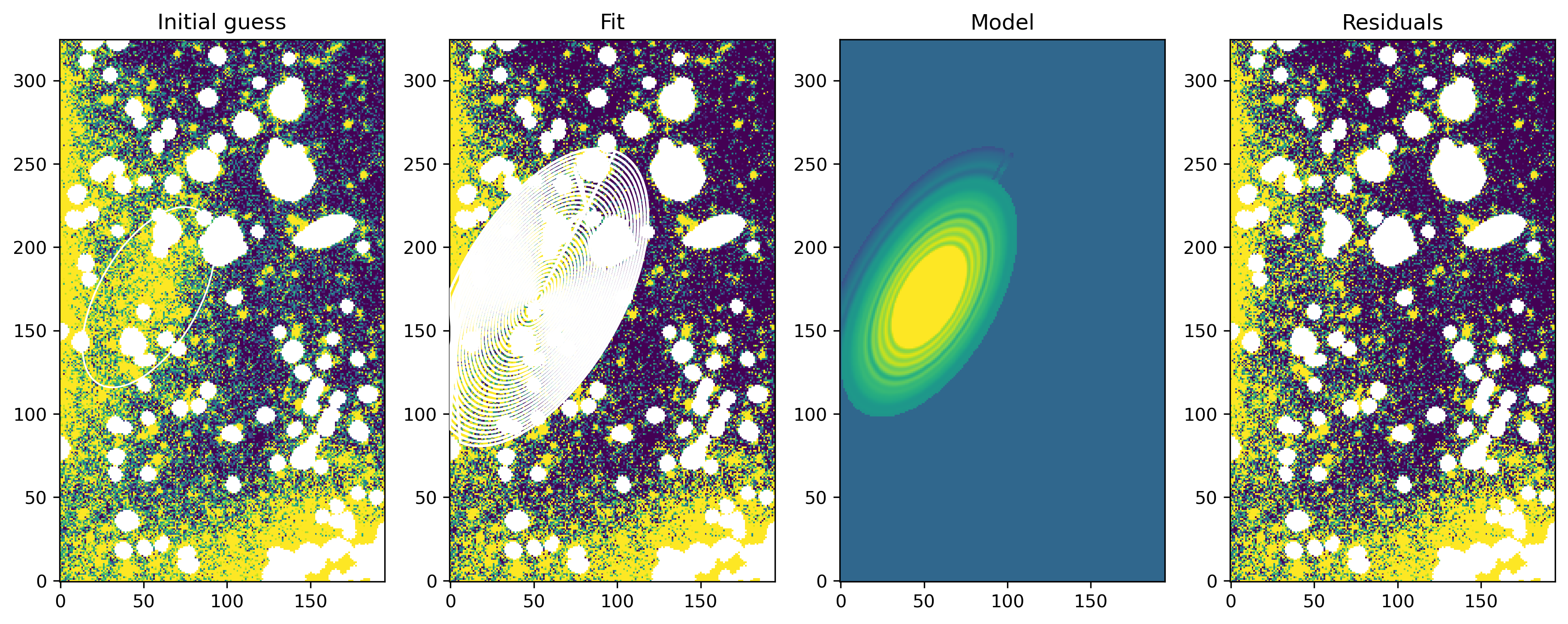}\\
    \includegraphics[width=.75\textwidth]{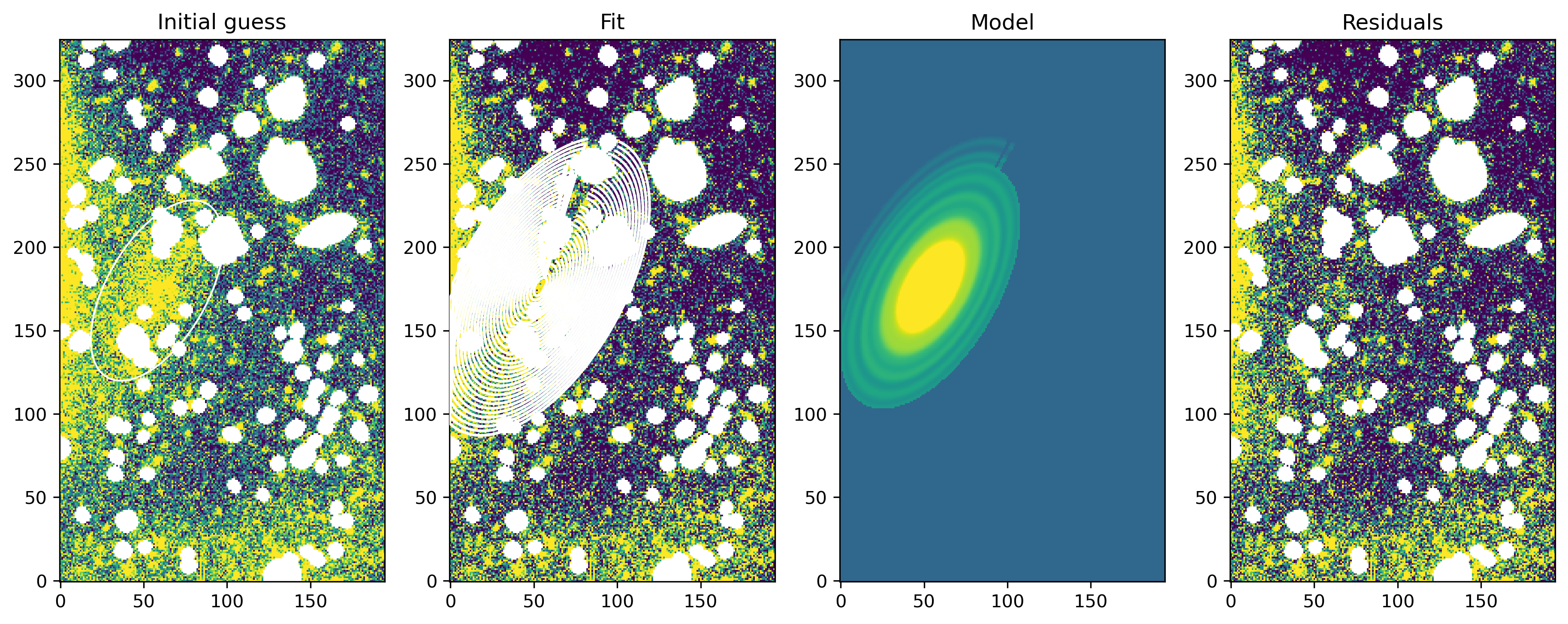}\\
    \caption{Modelling of the candidate ultra diffuse galaxy. Step-by-step results of our fitting routine using the \textit{Ellipse} function of \texttt{photutils.isophote}: a guess for the initial ellipse (\textit{left}); the fit out to a pre-determined radius (\textit{centre left}); the obtained elliptical model (\textit{centre right}); the residuals once having subtracted the model (\textit{right}), for the \textit{g}- (\textit{top}) and the \textit{r}- (\textit{bottom}) bands. Masked areas are marked in white. Each image has a binning of $2\times2$ and size of $\sim3\arcmin\times5\arcmin$. North is up and east to the left.}
    \label{fig:SB_UDG_Ellipse}
\end{figure*}

Figure~\ref{fig:scattered_light} shows the results of the scattered light removal procedure for the bright star BD+25~2039. In Fig.~\ref{fig:SB_UDG_Ellipse} is shown instead the elliptical modelling of the UDGC. Such procedure was also applied to the rest of DGCs.

\begin{figure*}[h!]
    \centering
    \includegraphics[width=0.4\textwidth]{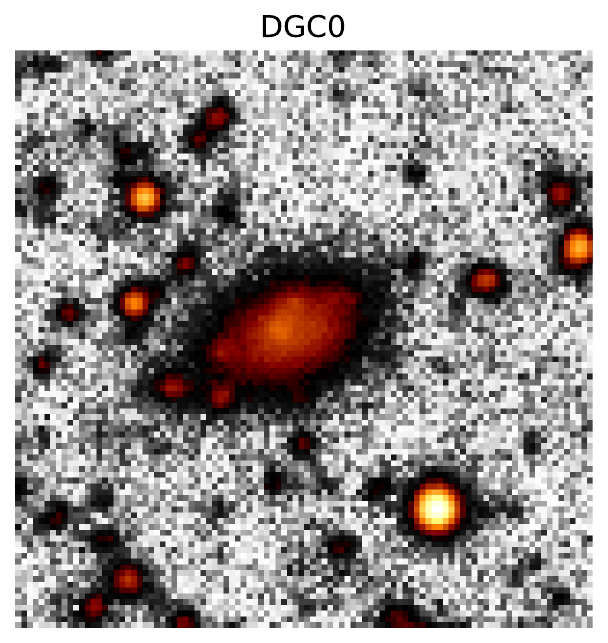}
    \includegraphics[width=0.4\textwidth]{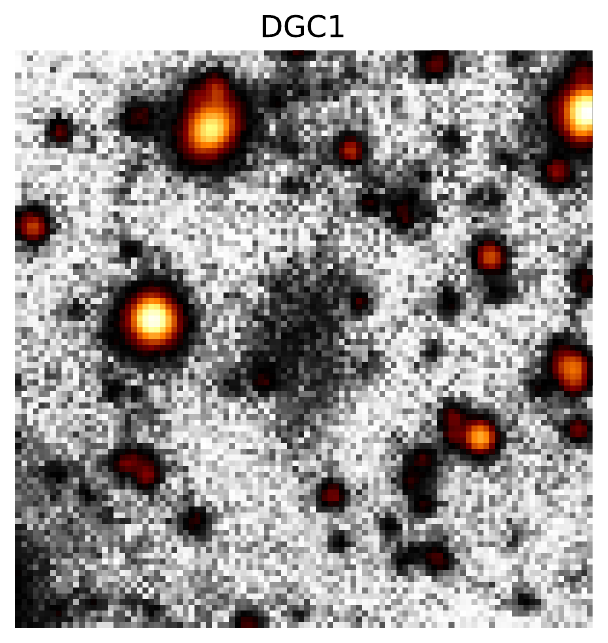}
    \includegraphics[width=0.4\textwidth]{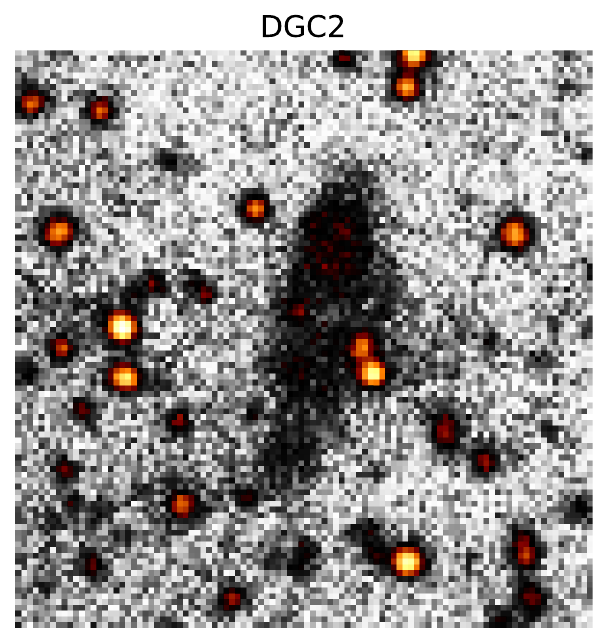}
    \includegraphics[width=0.4\textwidth]{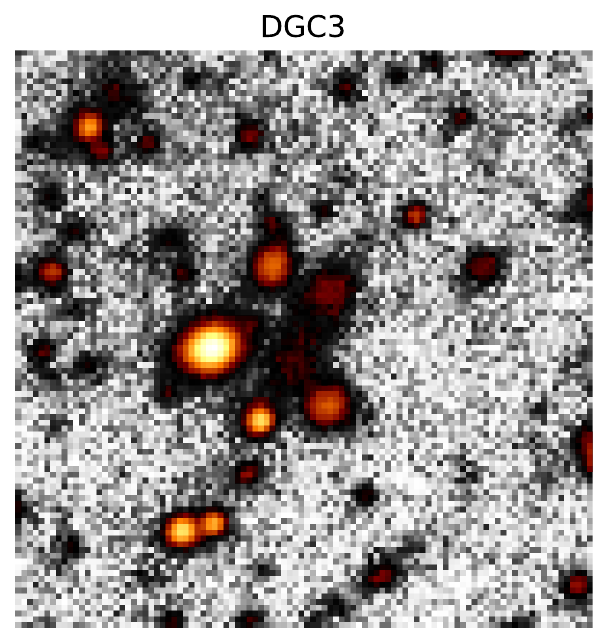}
    \includegraphics[width=0.4\textwidth]{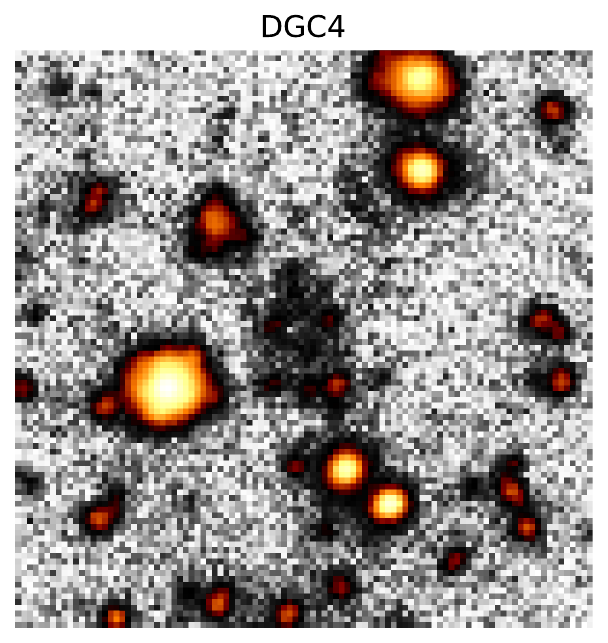}
    \includegraphics[width=0.4\textwidth]{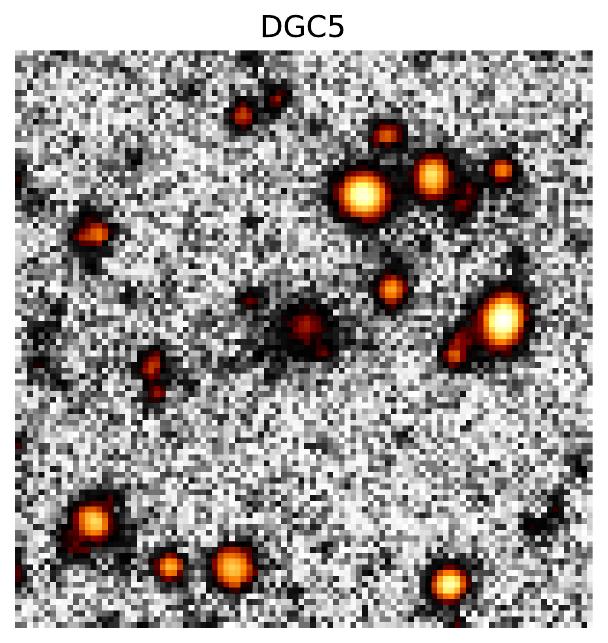}
    \caption{Cutouts of the co-added $(g+r)/2$ image around the candidate dwarf galaxies reported in this paper. Each image has a binning of $2\times2$ and a size of $45\arcsec\times45\arcsec$. North is up and east to the left.}
    \label{fig:lbc_DG}
\end{figure*}

\begin{figure*}[h!]
    \centering
    \includegraphics[width=0.4\textwidth]{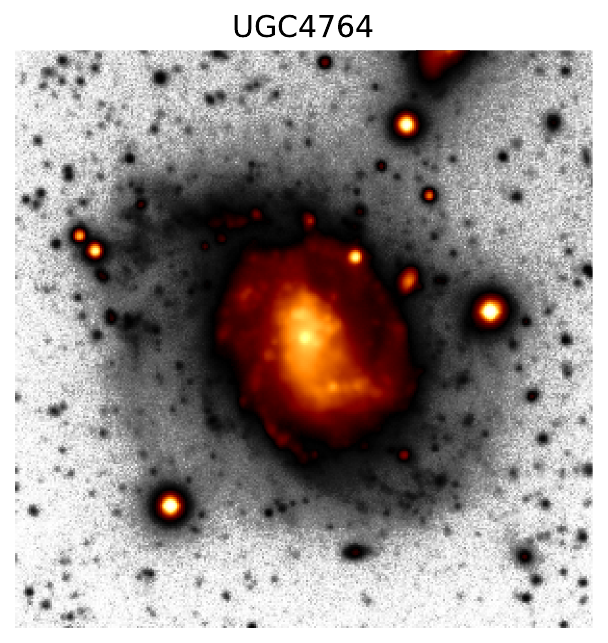}
    \includegraphics[width=0.4\textwidth]{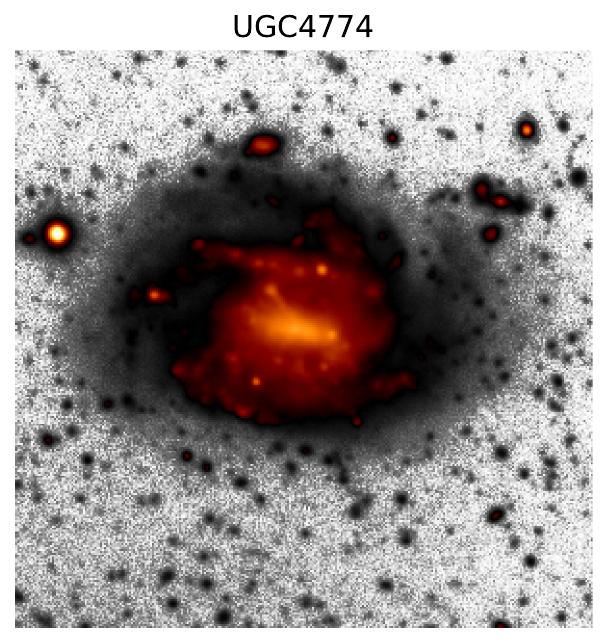}
    \includegraphics[width=0.4\textwidth]{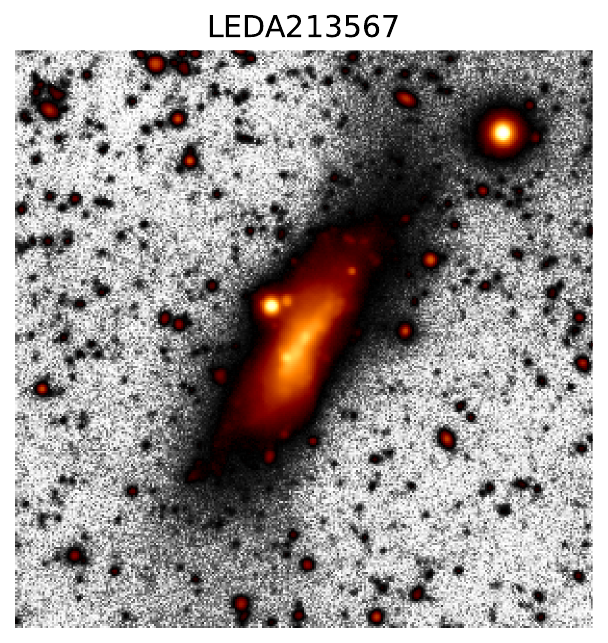}
    \includegraphics[width=0.4\textwidth]{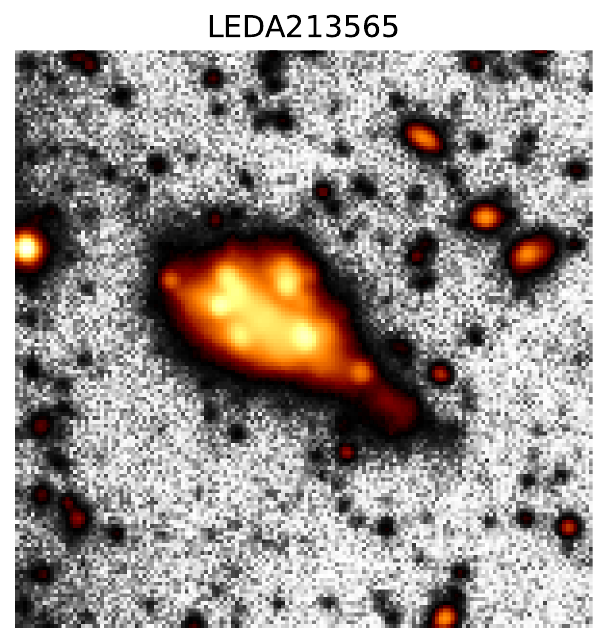}
    \includegraphics[width=0.4\textwidth]{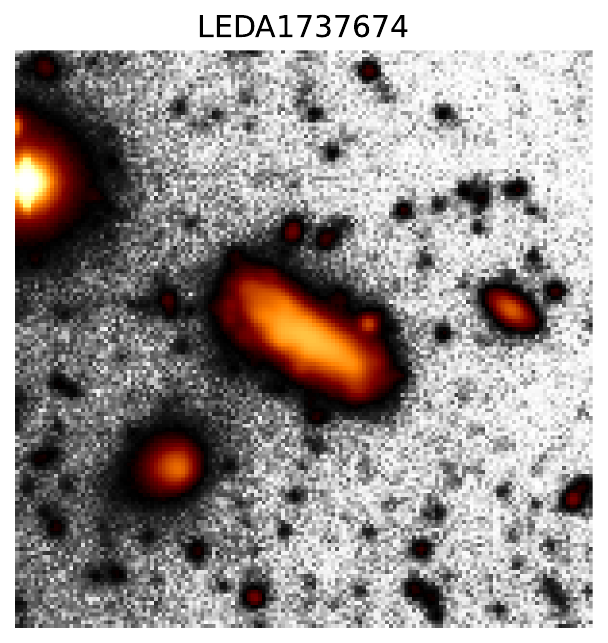}
    \includegraphics[width=0.4\textwidth]{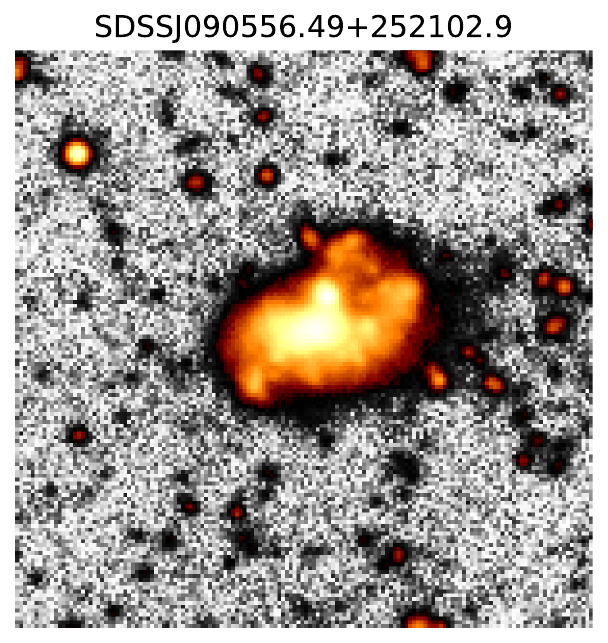}
    \caption{Cutouts of the co-added $(g+r)/2$ image around the rest of the galaxies belonging to the NGC~2750 system. Each image has a binning of $2\times2$ and size of $\sim2\arcmin\times2\arcmin$ (top three panels) or $\sim1\arcmin\times1\arcmin$ (all the others). North is up and east to the left.}
    \label{fig:lbc_glx}
\end{figure*}

\begin{figure*}
    \centering
    \includegraphics[width=\textwidth]{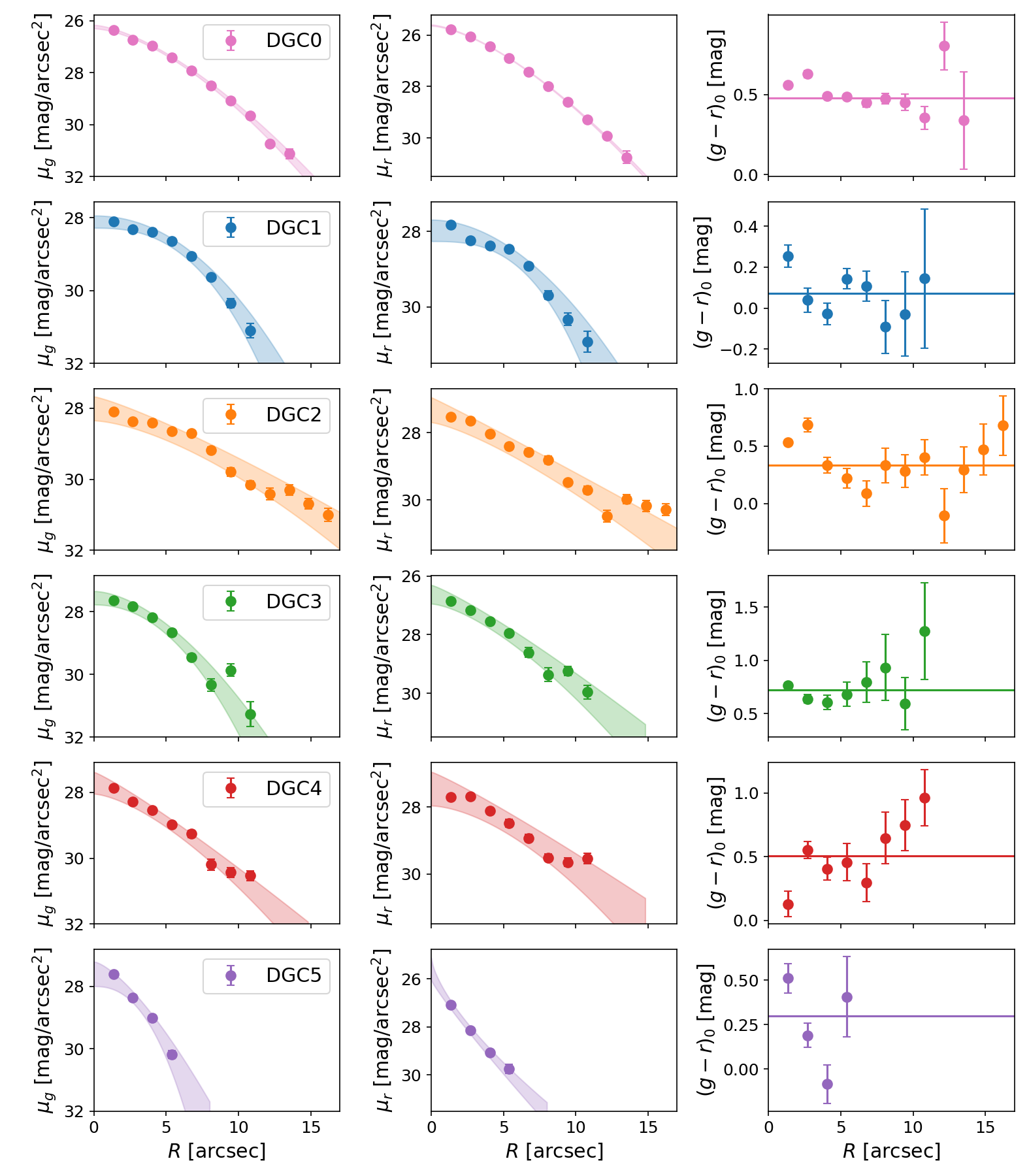}
    \caption{Radial surface brightness profiles in the \textit{g}- and \textit{r}-bands (left and middle columns) with the $1\sigma$ confidence interval of the best-fitting Sérsic profiles, together with the de-reddened colour profiles (right column) for the DGCs around NGC~2750, where the horizontal solid line represents the median colour value.}
    \label{fig:SBrad}
\end{figure*}

Finally, in Figs.~\ref{fig:lbc_DG} and \ref{fig:lbc_glx} we show the cutouts of all the galaxies (potentially) associated with NGC~2750, with panels having different binning and sizes. 
In particular, in Fig.~\ref{fig:SBrad} we show the radial surface brightness profiles in the \textit{g}- and \textit{r}-bands for the DGCs around NGC~2750.

\FloatBarrier
\clearpage

\end{appendix}

\end{document}